\documentclass[11pt,a4paper]{article}

\usepackage{jcappub}

\usepackage{epsfig}
\usepackage{url}
\usepackage{hyperref}

\usepackage{latexsym}
\usepackage{epsfig}
\usepackage{amssymb}
\usepackage{graphicx}
\usepackage{verbatim}

\newcounter{num}
\title{Tension in the Void: Cosmic Rulers Strain Inhomogeneous Cosmologies}

\author[a,c]{Miguel Zumalac\'arregui,}
\author[b]{Juan Garc\'ia-Bellido,}
\author[a]{Pilar Ruiz-Lapuente}
\affiliation[a]{Institut de Ciencies del Cosmos, Universitat de Barcelona IEEC-UB,\\ Marti i Franques 1, E-08028 Barcelona, Spain}
\affiliation[b]{Instituto de F\'isica Te\'orica UAM-CSIC, Universidad Aut\'onoma de Madrid,\\ Cantoblanco, 28049 Madrid, Spain}
\affiliation[c]{Institute of Theoretical Astrophysics, University of Oslo, 0315 Oslo, Norway}
\emailAdd{miguelzuma@icc.ub.edu}
\emailAdd{juan.garciabellido@uam.es}
\emailAdd{pilar@am.ub.es}

\arxivnumber{1201.2790}

\abstract{
New constraints on inhomogeneous Lema\^itre-Tolman-Bondi (LTB) models alternative to Dark Energy are presented, focusing on adiabatic profiles with space-independent Big Bang and baryon fraction. The Baryon Acoustic Oscillation (BAO) scale at early times is computed in terms of the asymptotic value and then projected to different redshifts by following the geodesics of the background metric. Additionally, a model-independent method to constraint the local expansion rate using a prior on supernovae luminosity is presented.
Cosmologies described by an adiabatic GBH matter profile with $\Omega_{\rm out}=1$ and $\Omega_{\rm out}\leq 1$ are investigated using a Markov Chain Monte Carlo analysis including the latest BAO data from the WiggleZ collaboration and the local expansion rate from the Hubble Space Telescope, together with Union-II type Ia supernovae data and the position and height of the Cosmic Microwave Background acoustic peaks. 
The addition of BAO data at higher redshifts increases considerably their constraining power and represents a new drawback for this type of models, yielding a value of the local density parameter $\Omega_{\rm in}\gtrsim 0.2$ which is 3$\sigma$ apart from the value $\Omega_{\rm in}\lesssim 0.15$ found using supernovae. The situation does not improve if the asymptotic flatness assumption is dropped, and a Bayesian analysis shows that constrained GBH models are ruled out at high confidence. We emphasize that these are purely geometric probes, that only recently have become sufficiently constraining to independently rule out the whole class of adiabatic LTB models.
}

\begin{document}

\maketitle

\section{Introduction}

Recent years have witnessed enormous advance in the quantitative understanding of cosmology and the establishment of a Standard Cosmological Model. Its construction is grounded on the general relativistic description of space-time with the usual Einstein equations. The Ansatz for the space-time is a spatially homogeneous and isotropic Friedmann-Robertson-Walker (FRW) metric, chosen to satisfy the generalized Copernican Principle, or Cosmological Principle.
In addition to the known particles (baryons, photons and neutrinos), two mysterious elements need to be added in order to account for all the observations. These give the name to the standard, $\Lambda$CDM model: Cold Dark Matter (CDM) plus a cosmological constant ($\Lambda$), the last one necessary to explain the dimming of distant supernovae \cite{Perlmutter:1998np,Riess:1998cb}. With the standard choice of the metric, the supernovae data imply that the universe is currently undergoing a phase of accelerated expansion.

The situation changes when the Cosmological Principle Hypothesis is dropped. As the supernovae we observe occur in our past lightcone, the changes in luminosity we interpret as time evolution might be due to spatial variations if the universe is not homogeneous. If the inhomogeneity represents an underdensity with a size comparable to the Hubble radius and our galaxy is located near its center, supernovae observations can be successfully accounted for without the introduction of new physics. This type of so-called ``large void models" can be described by a spherically symmetric Lema\^itre-Tolman-Bondi (LTB) metric, and have been studied as an alternative to the standard $\Lambda$CDM scenario \cite{Mustapha:1998jb,Celerier:1999hp,Tomita:2000jj, Alnes:2005rw,Vanderveld:2006rb,Garfinkle:2006sb,Enqvist:2007vb,Mattsson:2007tj,Sarkar:2007cx,
Zibin:2008vk,Clifton:2008hv,Moffat:2009bc,Celerier:2009sv,Vanderveld:2009eu,Regis:2010iq,Buchert:2011yu,Ellis:2011hk,Buchert:2011sx}.
Many different aspect of these alternative models have been considered over the past years, including observational constraints 
\cite{Enqvist:2006cg,Alexander:2007xx,GarciaBellido:2008nz,GarciaBellido:2008yq,Romano:2009mr,
February:2009pv,Quartin:2009xr,Moss:2010jx,Biswas:2010xm,Dunsby:2010ts,Nadathur:2010zm,Marra:2010pg,Zhang:2010fa,
Yoo:2010hi,Marra:2011ct,Bull:2011wi,Romano:2011tz,Wang:2011kj},
the growth of perturbations \cite{Zibin:2008vj,Clarkson:2009sc,Alonso:2010zv,Nishikawa:2012we,Alonso:2012ds,February:2012fp},
and the physics of the Cosmic Microwave Background (CMB) \cite{Moffat:2005yx,Alnes:2006pf,Caldwell:2007yu,GarciaBellido:2008gd,Clifton:2009kx,
Yoo:2010qy,Yoo:2010ad,Clarkson:2010ej,Moss:2011ze,Zibin:2011ma,Clifton:2011sn}. See reference \cite{Clarkson:2012bg} for a recent review.

If the time to Big Bang and the baryon-mater ratio are independent of the location, the LTB type of metric represents the gravitational growth of an adiabatic perturbation from an initially quasi-homogeneous state \cite{Zibin:2008vj}. Although a Gigaparsec-sized void is difficult to reconcile with the standard inflationary paradigm, it might still be possible through large non-perturbative inhomogeneities associated with the stochastic nature of the inflaton evolution~\cite{Linde:1994gy}.
The only philosophical problem associated with this type of models is the requirement of being located very close to the center in order to preserve the great degree of isotropy observed in the CMB \cite{Alnes:2006pf}, but ultimately, only cosmological observations can tell us on the geometry and distribution of the cosmos and our position in it, provided that this question is meaningful.

The aim of this work is to analyze LTB models in the light of the most recent cosmological data. The Hubble Space Telescope has made a precise  measurement of the local expansion rate \cite{Riess:2011yx} that challenges this type of void models, which typically require a low value to fit the CMB power spectrum \cite{Moss:2010jx,Biswas:2010xm}. The determination of the Hubble parameter relies on the calibration of distant supernovae using Cepheid variable stars and the subsequent fit to a fiducial $\Lambda$CDM model in the low redshift range. Although applying priors directly on $H_0$ is fine for homogeneous cosmologies, LTB universes can have a very different evolution in the relevant redshift range. Therefore, our analysis is based on the intrinsic Ia supernova luminosity instead of the value of the Hubble parameter.

Further restrictions on this model are obtained from the scale of Baryon Acoustic Oscillations (BAO) and its evolution in an inhomogeneous cosmology by including the most recent BAO data up to redshift $z=0.8$ provided by the WiggleZ collaboration \cite{Blake:2011en} and Carnero \textit{et al.} \cite{Carnero:2011pu}.
In a FRW cosmology, the BAO scale is space-independent and the constraints it yields arise from providing an independent measurement of cosmic distances relative to a \emph{standard ruler}, with an initial length determined by the physics of the early universe. Once baryons decouple, the dominant effect on the observed physical scale is to be stretched by the expansion of the universe. However, in the less symmetric LTB cosmology the initially constant BAO scale grows differently: the physical scale acquires an additional radial dependence and is stretched differently in the longitudinal and transverse direction, due to the different expansion rates.

Adding information about the BAO scale at higher redshift reduces considerably the room for its value in the early universe.
If the depth of the void is chosen to fit the supernovae luminosity distances, the inhomogeneous expansion produces a mismatch between the BAO scale at low and high redshift, posing a new problem for these models. Adding information on the CMB increases the discrepancies by restricting the initial acoustic scale, but the constraints are independent of the particular values, i.e, depend only on the {\it geometric} properties of the model regardless of the calibration of the standard rulers and candles. In particular, these constraints are independent of the primordial power spectrum, a critical assumption necessary to rule out large void models using the tension between the CMB and the local expansion rate \cite{Nadathur:2010zm} and are therefore complementary to these.

In section~2 we describe the general LTB void models, giving the corresponding Einstein-Friedmann equations, as well as the standard solution in absence of pressure. In a subsection we describe the adiabatic assumption, i.e. that the time since Big Bang is space independent, and thus the model only depends on a single function, the inhomogeneous matter profile $\Omega_M(r)$. This is chosen to have the the GBH parameterization, presented in a subsection.
In section~3 we study in detail the evolution of baryonic features in terms of free-falling trajectories of the background metric and compute the BAO observables.
In section~4 we describe the cosmological data used to analyze the model, including a method to use the supernova luminosity to constrain the local expansion rate.
In section~5 we analyze the results from the comparison of the model and the data and describe the tensions between the different datasets, as well as the result of different model comparison criteria.
Finally, in section~6 we give our conclusions, discussing the generality of the results and stating several modifications that might render the model viable.

\section{Lema\^itre-Tolman-Bondi Models}

The LTB model describes general spherically symmetric space-times and can be used as a toy model for describing voids in the universe \cite{Lemaitre:1933gd,Tolman:1934za,Bondi:1947av}. The starting point is the general metric
\begin{equation}\label{eq:metric}
ds^2 = - dt^2 + X^2(r,t)\,dr^2 + A^2(r,t)\,d\Omega^2\,,
\end{equation}
where $d\Omega^2 = d\theta^2 + \sin^2\theta d\phi^2$. Units in which $c=1$ will be assumed in the folowing. Assuming a spherically symmetric matter source with negligible pressure,
$T^\mu_{\ \nu} = - \rho_M(r,t)\,\delta^\mu_0\,\delta^0_\nu \,$,
the $(0,r)$ component of the Einstein equations, $G^t_{\ r} = 0$, sets the form of $X(r,t)$. The resulting cosmological metric becomes
\begin{equation}\label{ltb:metric}
ds^2 = - dt^2 + \frac{A'^2(r,t)}{1-k(r)}\,dr^2 + A^2(r,t)\,d\Omega^2\,,
\end{equation}
with an arbitrary function $k(r)$ playing the role of the spatial curvature parameter.
The other components of the Einstein equations read \cite{Enqvist:2006cg,Enqvist:2007vb,GarciaBellido:2008nz}
\begin{eqnarray} \label{eq:FRW1}
H_T^2 + 2H_T H_R + \frac{k(r)}{A^2} + \frac{k'(r)}{A A'} &=& 8\pi\,G\,\rho_M \,, \\
2{\dot H}_T + 3H_T^2 + \frac{k(r)}{A^2} &=& 0\,,
\end{eqnarray}
where dots and primes denote $\partial_t$ and $\partial_r$ respectively, 
and we have defined the transverse (i.e. in the angular direction) and radial Hubble rates as
\begin{equation}
 H_T\equiv \dot A / A,\;\; {\rm and} \;\;\; H_R\equiv \dot A' / A'\,.
\end{equation}
The reduced Hubble rate can be defined as usual: $H_{R/T} \equiv 100 h_{R/T} {\rm Km/Mpc/s}$. 
It is also useful to consider the normalized {\it shear}
\begin{equation}\label{shear}
\varepsilon \equiv \frac{H_T-H_R}{H_R+2H_T}\,,
\end{equation}
i.e. the difference between the radial and transverse expansion weighted by the total expansion \cite{GarciaBellido:2008yq}. This variable provides a local quanitfication of the departures with respect to homogeneous cosomologies, e.g. to characterize the growth of structure \cite{Alonso:2012ds}.

Integrating (\ref{eq:FRW1}) yields
\begin{equation}
H_T^2 = \frac{F(r)}{A^3} - \frac{k(r)}{A^2}\,,
\end{equation}
in terms of another arbitrary function $F(r)$. Substituting it into the first 
equation gives
\begin{equation}\label{F2rho}
\frac{F'(r)}{A'A^2(r,t)} = 8\pi\,G\,\rho_M(r,t)\,,
\end{equation}
where $\rho_M(r,t)$ is the physical matter density. Since $F(r)$ is
time-independent, one can choose $t=t_0$ and compute the integrated 
matter density in a comoving volume today, $V=4\pi r^3/3$, as
$\bar\rho(r) = \frac{1}{V}\int_0^r 4\pi r'^2dr'\,\rho_M(r',t_0)\,,$
and construct with it the ratio
$\Omega_M(r) \equiv \bar\rho(r)/\bar\rho_c(r)$, where 
$\bar\rho_c(r)=3H_0^2(r)/8\pi G$ is the critical density in that volume \cite{GarciaBellido:2008yq}.

\begin{figure}
\begin{center}
 \includegraphics[width=0.6\columnwidth]{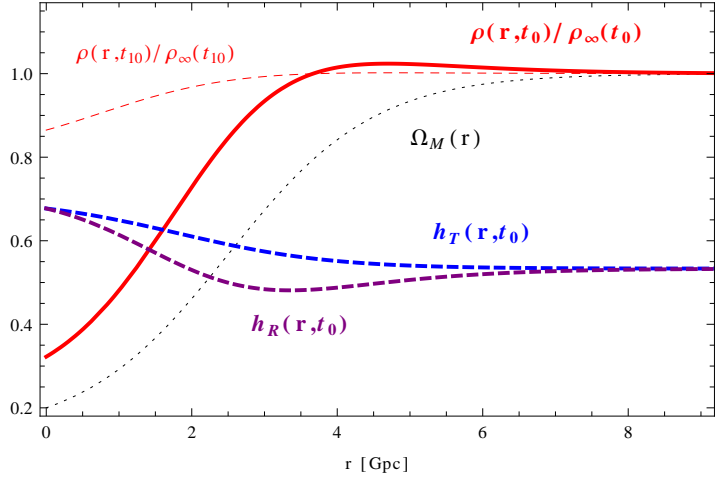}
\end{center}
\caption{Physical parameters in the LTB model. The density contrast at $t(z=0),t(z=10)$ (red) shows the evolution from an initially less inhomogeneous state, and differs from the function $\Omega_M(r)$ (black dotted). The expansion rates in the radial (purple dashed) and transverse (blue dashed) directions differ the most where the void is steeper. The profile shown has $R= 2.5, \Delta R=1$.
\label{physical}}
\end{figure}

The boundary condition functions $F(r)$ and $k(r)$ are specified by
the nature of the inhomogeneities through the local Hubble rate, the
integrated mass ratio and the local spatial curvature,
\begin{eqnarray}
F(r) &=& H_0^2(r)\,\Omega_M(r)\,A_0^3(r) = 8\pi G\int_0^r dr' r'^2 \rho_M(r',t_0)\,, \\
k(r) &=& H_0^2(r)\Big(\Omega_M(r)-1\Big)\,A_0^2(r) \,, \label{kdef}
\end{eqnarray}
where functions with subscripts $0$ correspond to present day values,
$A_0(r) \equiv A(r,t_0)$ and $H_0(r) \equiv H_T(r,t_0)$. With these definitions, the (position dependent) transversal Hubble rate can be written 
as~\cite{Enqvist:2006cg,Enqvist:2007vb}
\begin{equation}\label{eq:hubblerate}
H_T^2(r,t) = H_0^2(r)\left[\Omega_M(r)\left(\frac{A_0(r)}{A(r,t)}\right)^3 +
(1-\Omega_M(r))\left(\frac{A_0(r)}{A(r,t)}\right)^2\right]\,,
\end{equation}
and we fix the gauge by setting $A_0(r)=r$. For fixed $r$ and $\Omega_M<0$ the above expression
is equivalent to the Friedmann equation, and has an exact parametric solution in terms of the variable $\eta$:
\begin{eqnarray}
A(r,t) &=& \frac{\Omega_M(r)}{2[1-\Omega_M(r)]} [\cosh(\eta) - 1] A_0(r)\,, \label{eq:Aeta} \\
H_0(r) t &=& \frac{\Omega_M(r)}{2[1-\Omega_M(r)]^{3/2}} [\sinh(\eta) - \eta]\,. \label{eq:timeeta}
\end{eqnarray}
Very good approximate solutions can also be found by
Taylor expanding around an Einstein de Sitter solution \cite{GarciaBellido:2008nz}.

In addition to the solution of Einstein Equations (\ref{eq:Aeta},\ref{eq:timeeta}) it is necessary to obtain the coordinates on the lightcone as a function of redshift. For light traveling along radial null geodesics, $ds^2=d\Omega^2=0$ yields
\begin{equation}
\frac{dt}{dr} = \mp \frac{A'(r,t)}{\sqrt{1-k(r)}}\,,
\end{equation}
which, together with the redshift equation \cite{Enqvist:2006cg,Bondi:1947av},
\begin{equation}\label{eq:null}
\frac{d\log(1+z)}{dr} = \pm \frac{\dot A'(r,t)}{\sqrt{1-k(r)}}\,,
\end{equation}
allows us to write a parametric set of differential equations, with $N=\log(1+z)$ being the effective number of e-folds
before the present time,
\begin{eqnarray}\label{eq:lightrayt}
&&\frac{dt}{dN} = - \frac{A'(r,t)}{\dot A'(r,t)} \,,\\ \label{eq:lightrayr}
&&\frac{dr}{ dN} = \pm \frac{\sqrt{1-k(r)}}{\dot A'(r,t)}\,,
\end{eqnarray}
where the equations are integrated with the initial condition $r(0)=0$, $t(0)$ obtained from (\ref{eq:Aeta},\ref{eq:timeeta}) for $r=0,A=A_0$.

The angular diameter distance is given by the $d\Omega$ element of the metric evaluated on the lighctone, and is related to the luminosity distance by the redshift due to photon redshift and time dilation
\begin{eqnarray}
&& D_A(z) = A(r(z),t(z))\,,\\
&& D_L(z) = (1+z)^2A(r(z),t(z))\,.
\end{eqnarray}

The dynamics of the LTB metric in the only-matter approximation discussed above are fully specified by the two functions $\Omega_{M}(r)$, $H_0(r)$ independently of the type of matter present, as long as it exerts no pressure. But by dropping the symmetries of the FRW model a spherically symmetric but inhomogeneous mixture of baryonic and dark matter can be accommodated. A possible parameterization in terms of the total matter density would be
\begin{equation}\label{baryonfraction}
 f_b(r) \equiv \frac{\rho_b(r,t)}{\rho_m(r,t)} \,,
\end{equation}
where there is no time dependence because the energy density of baryons and dark matter evolves identically at late times.

Before explaining the choice of matter profile and the physical restrictions on the model, let us briefly summarize the approximations used throughout this work
\begin{itemize}
\item Spherical symmetry as given by (\ref{ltb:metric}) and perfectly central location of our galaxy at $r=0,\,t=t_0$ as initial conditions for the lightcone integration (\ref{eq:lightrayt}).
\item Radiation energy and pressure neglected as a source of the expansion (\ref{eq:hubblerate}).
\item Early time and large radius FRW limit of the model, necessary to compute the BAO scale (Section \ref{baoltb}) and the relative locations of the CMB peaks (Section \ref{cmb-section}).
\item Perturbations of the LTB metric neglected. The evolution of the BAO scale from early times is studied by analyzing the geodesics of the background metric (\ref{ltb:metric}) (Section \ref{baoltb}).
\end{itemize}

\subsection{The constrained (adiabatic) GBH model}\label{section_cgbh}

General LTB models are uniquely specified by the two functions $k(r)$ and $F(r)$ or equivalently by $H_0(r)$ and $\Omega_M(r)$, but to test them against data it is necessary to parameterize the functions, so that a finite dimensional space is analyzed. In this paper we will use the GBH model \cite{GarciaBellido:2008nz} to describe the matter profile in terms of a reduced number of parameters. In addition to the choice for the free function $\Omega_M(r)$, we further impose that the time to Big Bang is space-independent
\begin{equation}
t_{\rm BB}(r) = H_0(r)^{-1}\left({1\over \sqrt{\Omega_K(r)}}\sqrt{1 + {\Omega_M(r)\over
\Omega_K(r)}} - {\Omega_M(r)\over\sqrt{\Omega_K^3(r)}}\ {\rm sinh}^{-1}
\sqrt{\Omega_K(r)\over\Omega_M(r)}\right) = t_0 \,,\label{TBB2}
\end{equation}
where $\Omega_K(r) = 1 - \Omega_M(r)$. The above expression can be obtained from integration of (\ref{eq:hubblerate}) \cite{Enqvist:2006cg,GarciaBellido:2008nz} or by solving for $A=A_0(r)$ in (\ref{eq:Aeta},\ref{eq:timeeta}). This condition reduces the functional freedom associated to $H_0(r)$ to a single normalization constant $H_0$, that is related to the overall age of the universe.

Additionally, we require that there are no large scale baryonic isocurvature modes, i.e. the baryon fraction (\ref{baryonfraction}) is constant. This type of voids can be regarded as the gravitational collapse of a large scale, adiabatic and spherically symmetric perturbation which has a small amplitude at early times. Its adiabatic nature is related to the fact that there is only one functional degree of freedom that sets the shape of the remaining free functions ($\Omega_M(r)$ in our case, which in turn fixes $H_0(r)$ and the baryon fraction).

The above conditions give a relation between $H_0(r)$, $\Omega_M(r)$ and $f_b(r)$, and hence constrain the models to one free function. Our chosen model is thus given by
\begin{eqnarray}
\hspace{-2cm}&&\Omega_M(r) = \Omega_{\rm out} + \Big(\Omega_{\rm in} - 
\Omega_{\rm out}\Big)
\left({1 - \tanh[(r - R)/2\Delta R]\over1 + \tanh[R/2\Delta R]}\right) \\
\hspace{-2cm}&&H_0(r) = H_0\left[{1\over \Omega_K(r)} -
{\Omega_M(r)\over\sqrt{\Omega_K^3(r)}}\ {\rm sinh}^{-1}
\sqrt{\Omega_K(r)\over\Omega_M(r)}\right] =
H_0 \sum_{n=0}^\infty {2[\Omega_K(r)]^n\over(2n+1)(2n+3)}\,,\label{H20} \\
\hspace{-2cm}&&\, f_b(r) = f_b = \mbox{constant}
\end{eqnarray}
where the second equation follows from (\ref{TBB2}), and the third by demanding constant baryon to matter ratio. This paremeterization was introduced to lower the shear around the void wall (e.g. with respect to Gaussian profiles) and allow an unified description of cuspy and flat central regions \cite{GarciaBellido:2008nz}. Each void model is specified by the following parameters:
\begin{itemize}
 \item $\Omega_{\rm in}$: Matter/curvature fraction at the center of the void (equations \ref{F2rho}-\ref{kdef}). As deeper voids produce more fictitious acceleration, this parameter plays a major role in the constraints presented in Section \ref{results}.
\item $\Omega_{\rm out}$: Asymptotic ($r\to\infty$) matter/curvature fraction in which the inhomogeneous region is embedded.
Two possibilities will be considered separately depending on the asymptotic curvature of the universe: 
\subitem -\textbf{CGBH:} Flat $\Omega_{\rm out}=1$, as suggested by inflationary physics.
\subitem -\textbf{OCGBH:} Open $\Omega_{\rm out} \leq 1$, which allows a better fit to the CMB.
\item $\Delta R$: Slope of the inhomogeneity. Smaller values of $\Delta R$ produce steeper profiles and increase the shear (\ref{shear}).
\item $R$: Shape of the void. $\Delta R \ll R$ describes an inhomogeneity with a central plateau of approximately constant density, while $\Delta R \gg R$ produces a cuspy central region.
\item $H_0$: Expansion rate normalization that determines the Big Bang time (\ref{TBB2}).
\item $f_b$: Baryon fraction over the total matter content. Its value affects the pre-recombination physics and determines the value of the BAO scale and CMB peak locations.
\end{itemize}

The choice of the constrained model is important because, in our gauge, void models with an {\it inhomogenous} Big Bang would contain a mixture of growing and decaying modes, and consequently the void would not disappear at early times, making them incompatible with the Standard Big Bang scenario~\cite{Zibin:2008vj}. By restricting ourselves to adiabatic LTB models the central void is reduced to an insignificant perturbation in an otherwise homogeneous universe described by an FRW metric, both at large distances \emph{and} early times. This requirement, together with the condition of constant baryon fraction, ensures the space-independence of the early BAO scale, which is a key part of the present analysis.

\section{The Baryon Acoustic Scale in LTB universes} \label{baoltb}

Inhomogeneous cosmologies stretch the BAO scale differently than their homogeneous cousins. There are three potential effects to be addressed when computing the BAO scale in LTB models at the background level:
\begin{enumerate}
\item Inhomogeneous expansion: The matter distribution will source the expansion of the universe in a position dependent way. Therefore, there will be a radial dependence of the physical scale in addition to the time dependence.
\item Anisotropic expansion: In general the expansion rate in the radial and transverse direction will be different $H_T\neq H_R$, resulting in two different BAO scales $l^T\neq l^R $, as seen by a central observer.\footnote{The relation between this effect and the Alcock-Paczynski test will be discussed in Section \ref{ap-test}.}
\item Radial coordinate drift: Displacements in the radial direction are not a symmetry of the LTB metric and the free-falling baryon features are not ensured to remain at constant $r$. In Section \ref{freefallbao} we show that this effect does not occur for timelike geodesics.
\end{enumerate}
Possible effects from higher order corrections will be discussed at the end of Section \ref{nextorder}.

\begin{figure}
\begin{center}
 \includegraphics[width=1.0\columnwidth]{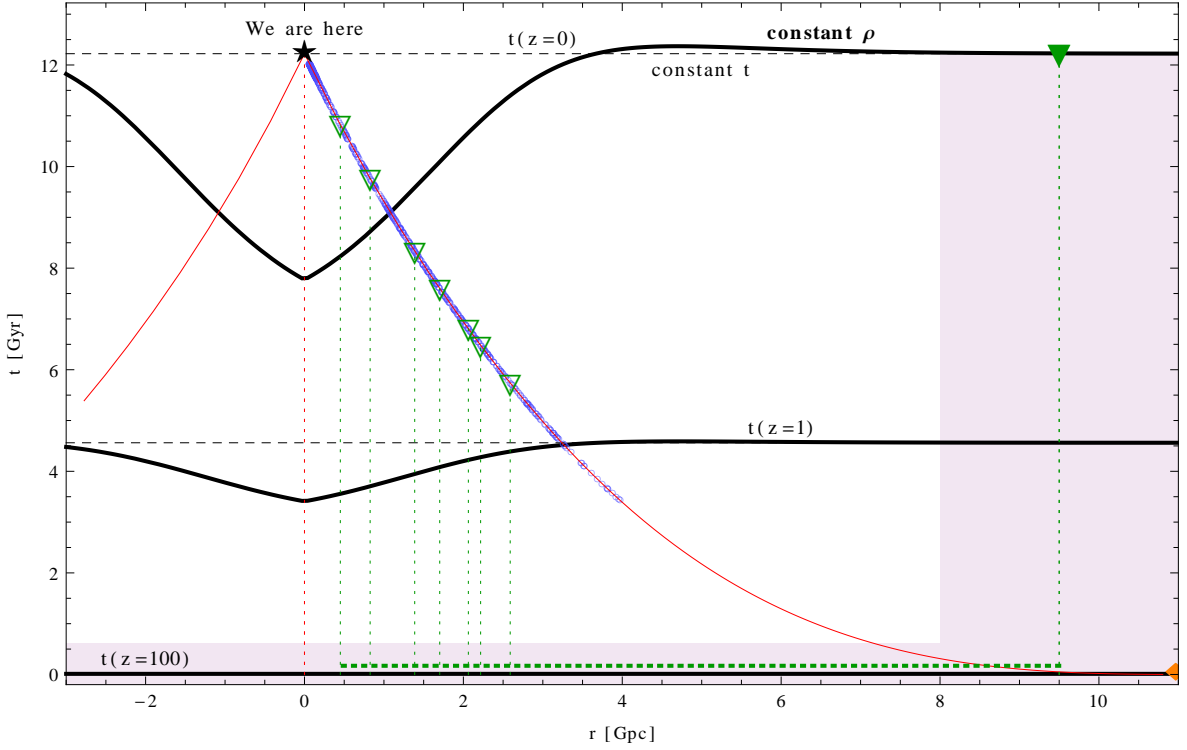}
\end{center}
\caption{The LTB model with space-independent Big Bang. Black lines represent hypersurfaces of constant time $t=t(z)$ for $z=0,1,100$ (dashed) and constant density $\rho(r,t)=\rho(r_\infty,t(z))$ (thick, continuous).
The filled areas show schematically the regions in which the void can be considered homogeneous (early times and large radii). Red constant lines represent our lightcone, where the coordinates of SNe (blue circles on the left) and BAO (green triangles on the right) observations has been added. 
Vertical dotted lines correspond to the geodesic worldlines of our galaxy (red) and the BAO fiducial locations (green). The physical BAO scale at different $z$ is obtained from the asymptotic value (represented by the filled green triangle), extrapolated to early times, for which the universe is approximately homogeneous (horizontal green dotted line) and evaluated at the lightcone coordinates using the LTB metric (see Section \ref{baoltb} for the details).
\label{baodiagram}}
\end{figure}

Our approach to predict the BAO scale in LTB models with space-independent Big Bang time relies on the homogeneity properties of the metric at large radius (i.e. the profile flattens) and early times (i.e. the Big Bang time and the baryon fraction are independent of the position). Relaxing this assumptions would require a more careful treatment which goes beyond the scope of this work. We first analyze the evolution of the BAO scale in the inhomogeneous cosmology by following the geodesics of the LTB metric (Section \ref{freefallbao}).
Afterwards, the asymptotic physical scale computed in the limit $r\gg R,\, \Delta R$ using the fitting formulae is extrapolated to a suitable early time $t_e$ at which the void is just a negligible perturbation. The obtained value can be then projected to the coordinates of observation $r(z),t(z)$ using the previous results (Section \ref{earlybao}). Finally, the physical scale is related to the observed quantity $d_z$ quoted by the galaxy surveys (Section \ref{obsbao}). The procedure is sketched in Figure \ref{baodiagram}.

The derivation we are presenting avoids using certain concepts that might be equivocal when used for inhomogeneous cosmologies. In particular, we will only use redshift as a coordinate on the lightcone or in the asymptotic region when the FRW limit can be applied. We will also avoid the term ``comoving'' and will refer to ``coordinate'' instead, as well as pay special attention to distinguishing physical distances and relative coordinate separations.

\subsection{Free-falling scales in the LTB metric}\label{freefallbao}

The propagation of sound waves in the baryon-photon fluid present in the early, expanding, universe leaves an imprint at a characteristic length that will be observable in the late universe as a peak in the correlation function of galaxies \cite{Eisenstein:1997ik,Eisenstein:2005su}. When the universe becomes neutral, baryon-photon interactions render effectively zero and the baryonic overdensities start behaving as free-falling test bodies. We can therefore analyze the relative separation of the initial baryon clumps and the galaxies they will form by following the geodesics of the LTB metric: $x^\mu(\tau)=\{t(\tau),r(\tau),\theta(\tau),\phi(\tau)\}$, where $\dot x^\mu=\frac{d x^\mu}{d\tau}$. The BAO scale can be traced simply by following two nearby trajectories with an initial separation equal to the baryon acoustic scale at a sufficiently early time.

The transverse evolution is the simplest. Since rotations are an isometry of the LTB metric, the momentum in the angular directions is conserved and trajectories with $\dot\phi,\dot\theta=0$ initialy will remain at constant angular coordinates. This can be readily seen from the geodesic equation for the $\theta$ coordinate
\begin{equation}
 \ddot \theta + 2\frac{A'}{A} \dot r \dot \theta + 2 \frac{\dot A}{A}\dot t \dot \theta =0\,,
\end{equation}
for which $\theta(\tau) =\theta_0$ is a solution. Its stability follows by demanding timelike, slow geodesics for which $\dot t \gg \dot r$ and noting that the second term is positive in an expanding universe ($\dot A>0$, $H_T>0$) and acts as a friction against the angular velocity $\dot\theta$. Therefore, initial angular separation $\Delta \theta$ is conserved in coordinate space and the associated, transverse physical scale can be obtained integrating the angular element of the metric $l^T_{\rm phys}=A(r,t)\Delta \theta$.

Since shifts in the radial direction are not an isometry of the LTB metric, $\dot r$ is not automatically conserved and the determination of the radial acoustic scale requires a more careful treatment. The geodesic equations for a trajectory with $\dot\phi,\dot\theta=0$ are
\begin{equation}\label{geodesict}
 \ddot t + \frac{A' \dot A'}{1-k(r)}\dot r^2 = 0 \,,
\end{equation}
\begin{equation}\label{geodesicr}
 \ddot r + \left[ \frac{k'(r)}{2(1-k(r))} + \frac{A''}{A'} \right]\dot r^2 + 2\frac{\dot A'}{A'}\dot t \dot r = 0 \,.
\end{equation}
Similarly to the angular case, a particle initially at rest at some early time $t_e$, $\dot r (t_e)=0$, will remain at constant radial coordinate location $r(\tau)=r(t_e)$.  
Timelike trajectories with $\dot t \gg \dot r$ are again stable in an expanding universe due to the longitudinal Hubble friction term $\dot A'/A'>0$, and geodesics will remain at constant coordinate separations at different cosmic epochs.\footnote{The situation would considerably change if $\Gamma^r_{tt}$ was different from zero in the LTB metric. It might be as well possible to devise profiles for which the first term in (\ref{geodesicr}) overcomes the second for sufficiently rapid geodesics, i.e. high $\dot r/\dot t$, but this is not the case for the models under study.} The physical distance in the $r$ direction can be obtained simply by integration using the radial element of the metric $l^R_{\rm phys}=\int \sqrt{g_{rr}}d r \approx A'/\sqrt{1-k} \Delta r$.

To summarize, since coordinate locations are conserved in geodesic evolution, we can provide the following relation for freely falling, physical scales in the LTB metric at different times $t,t_e$ in the transverse and longitudinal directions
\begin{equation}\label{Tconvert}
l^T_{\rm phys}(r,t)=\frac{A(r,t)}{A(r,t_e)} l^T_{\rm phys}(r,t_e)\,,
\end{equation}
\begin{equation}\label{Lconvert}
l^R_{\rm phys}(r,t)=\frac{A'(r,t)}{A'(r,t_e)} l^R_{\rm phys}(r,t_e)\,.
\end{equation}

\subsubsection{BAO scale evolution beyond zero order}\label{nextorder}

The above analysis so far has dealt with the differences between homogeneous and inhomogeneous models at the zero order level. 
In a FRW universe, the BAO scale is constant in coordinate space also in first order perturbation theory. This is due to the lack of scale dependence of sub horizon perturbations at late times $\delta_k(t)=D(t)\delta_k(t_0)$ \cite{Dodelson:2003ft}, which preserves the shape of the power spectrum and the two point correlation function (both are related by a Fourier transform).
The first corrections come through nonlinear effects, which act reducing the BAO scale at the few percent level, due to the attraction between the initial perturbation and the baryon clumps at the characteristic length \cite{Crocce:2007dt,Smith:2007gi}. These effects will eventually become relevant as the precision of surveys and the reconstruction techniques improve.

In inhomogeneous universes the situation is different, since the lack of symmetry produces the failure of the perturbation decomposition and the treatment of scalar perturbations (in the metric and matter density) have to be considered together with vector and tensor perturbations \cite{Zibin:2008vj,Clarkson:2009sc}. The vorticity (vector) component is subdominant because the LTB metric is rotationally invariant, but the scalar potential is sourced at linear order by a term proportional to the background shear traced with the tensor perturbations. Fortunately, for the GBH profiles considered here, the background shear is below 5\% and these contributions will be subdominant with respect to the much larger effect of the inhomogeneous expansion (see Figure \ref{baorescaling} and Section \ref{obsbao}). February {\it et al.} recently presented the numerical computation of the BAO scale in LTB models within linear theory, and considering only scalar perturbations \cite{February:2012fp}. They found a shift on the BAO scale at the percent level, which is nevertheless considerably smaller than the departures induced by the inhomogeneous and anisotropic expansion discussed above.

Further support for the assumption of a constant BAO scale in coordinate space is provided by N-body numerical studies. Alonso et al. run simulations in which inhomogeneous matter profiles are implemented through an initial underdensity of Gpc size. Their results show that the (local) matter density contrast grows with the scale factor in a way analogous to that of an open universe with a value of the matter density $\Omega_M(r)$ corresponding to the appropriate location $r$ \cite{Alonso:2010zv}, showing an effective decoupling between the small scale clustering and the evolution of the void. Corrections from the large scale inhomogeneity are proportional to the local shear weighted by a factor $\mathcal O(1)$ \cite{Alonso:2012ds}, and are hence small for the profiles allowed by observations.

\subsection{The physical BAO scale at early times and on the lightcone}\label{earlybao}

The solutions of the LTB metric with space-independent Big Bang represent an inhomogeneity that grows due to gravitaional instability out of a very homogeneous state. For typical voids at $t(z=100)$ the physical density contrast $\rho_m(r,t)/\rho_m(r_\infty,t)$ is of order 1\%, while at $t(z=1000)$ it shrinks to $\sim 0.1\%$. As the baryionic features develop between $t\sim 0$ and $t\sim t(z=1000)$, it is a good approximation to consider that the physics responsible for recombination and the origin of the baryon acoustic scale are indistinguishable from their counterparts in homogeneous cosmologies.
It will be therefore assumed that for models with space-independent Big Bang the physical BAO scale is isotropic and coordinate independent at early times on constant time hypersurfaces.\footnote{A more physical criterion would be to consider constant density hypersurfaces. Although the difference is of order $0.1\%$ in models with space-independent Big Bang, it might render helpful to generalize the treatment of BAO for profiles with general $H_0(r)$.}
\begin{equation}\label{earlyconvert}
 l_{BAO}(r(z),t_e)\approx l_{BAO}(r_\infty,t_e) \,.
\end{equation}

The early time BAO scale can be obtained from the asymptotic value at different times using equation (\ref{Tconvert}) or (\ref{Lconvert}). In the $r\to\infty$ limit the universe is indistinguishable from a FRW cosmology, and we can compute the BAO scale using the fitting formulae provided by Eisenstein and Hu \cite{Eisenstein:1997ik} in terms of the asymptotic values of the matter density and baryon fraction.
These effective values are obtained by projecting the LTB parameters on the lightcone at a very high redshift $z_e\approx 100$, for which 1) $H_T\approx H_R$ and the universe is approximately homogeneous on a constant $t=t(z_e)$ hypersurface, 2) the point $r(z_e)$ is away from the inhomogeneous region and 3) the radiation contribution is still negligible. These values are given by 
\begin{equation}
\Omega^{\rm eff}_m = \frac{\rho(r_\infty,t_0)}{3H_T ^2(r_\infty)} \approx \Omega_{\rm out}\,,
\end{equation}
\begin{equation}
\Omega^{\rm eff}_b\approx f_b \, \Omega_{\rm out}\,,
\end{equation}
\begin{equation}
 H^{\rm eff}_0= \frac{2 H_T(z_e) + H_R(z_e)}{3\sqrt{\Omega^{\rm eff}_m(1+z_e)^{3} + (1-\Omega^{\rm eff}_m)(1+z_e)^{2}}}\,.
\end{equation}
Here $H_0^{\rm eff}$ can be understood as ``rewinding'' the LTB value of the average expansion rate $2H_T(z_e)/3 + H_R(z_e)/3$ using the FRW asymptotic value of $\Omega_m$.

\begin{figure}
\begin{center}
 \includegraphics[width=0.6\columnwidth]{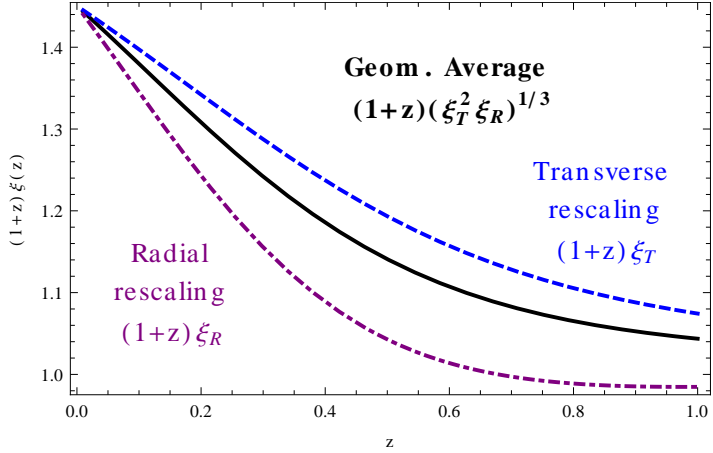}
\end{center}
\caption{Effects on the LTB metric on the BAO scale with respect to their FRW analogues. The difference between the transverse (blue, dashed) and longitudinal (purple, dot-dashed) factors accounts for the anisotropy of the scales in the angular and transverse direction, respectively. The geometric averaged rescaling factor (black solid) is used for volumetric BAO determinations through the quantity $d_z^{\rm LTB} = (1+z)\xi(z) \, d_z^{\rm FRW}$ (\ref{baoeq2}). Note that all three curves coincide in $r=0$, since the void is locally isotropic at the center.
\label{baorescaling}}
\end{figure}

The fitting formulae give a comoving scale in a FRW universe which coincides with the physical value at $t=t_0$ for the usual definition of the scale factor in the asymptotic FRW metric $a(t_0)=a_0=1$. The corresponding scale at $t_0$ is valid in the limit $r\to \infty$, but can be related to the radial and transverse physical scales using (\ref{Tconvert},\ref{Lconvert},\ref{earlyconvert}). For a point located on the past lightcone of the central observer, the values are
\begin{equation}\label{lT}
 l^T_{BAO}(z) \equiv \xi_T(z) l_{BAO}(r_\infty,t_0)
 = \frac{A(r(z),t(z))}{A(r(z),t_e)} \frac{A(r_\infty,t_e)}{A(r_\infty,t_0)} l_{BAO}(r_\infty,t_0) \,,
\end{equation}
\begin{equation}\label{lL}
 l^R_{BAO}(z) \equiv \xi_R(z) l_{BAO}(r_\infty,t_0)
= \frac{A'(r(z),t(z))}{A'(r(z),t_e)} \frac{A'(r_\infty,t_e)}{A'(r_\infty,t_0)} l_{BAO}(r_\infty,t_0) \,.
\end{equation}
The first equalities define a transversal and longitudinal rescaling factors (see Figure \ref{baorescaling}), which reduce to $(1+z)^{-1}$ in the homogeneous limit.

\subsection{Comparison with the observed BAO scale}\label{obsbao}

The BAO scale can be extracted from the galaxy correlation function measured in galaxy surveys. What is actually observed is a combination of the angular correlation $\theta_{BAO}$ and the correlation in redshift space $\Delta z_{BAO}$ \cite{Bassett:2009mm}. In order to compare the models with observations, we need to relate the isotropized correlation measured by the surveys
\begin{equation}\label{deltaz}
 d_z=\left(\theta_{BAO}^2 \frac{\Delta z_{BAO}}{z}\right)^{1/3}\,,
\end{equation}
to the physical scales computed in the previous section.\footnote{This section follows section 4.6.3 of Biswas et al. \cite{Biswas:2010xm}. Our result (\ref{baoeq1}) has the same form as their equation (4.48) after several coefficients cancel out. However, their computation assumes the BAO scale to be given by the local values of $\Omega_M(r),\Omega_B(r),H_0(r)$ instead of obtaining them from the asymptotic FRW value and the factor $(1+z_{rec})$ is taken to be given by the volume element comparison on the worldline of constant $r=r(z)$ instead of by the asymptotic value $r\to\infty$.}

The angular correlation can be readily obtained from the definition of the angular diameter distance as the ratio between a known (transverse) length and the angle it subtends
\begin{equation}\label{dtheta}
\theta_{BAO} = \frac{l^T_{BAO}(z)}{D_A(z)}\,.
\end{equation}
The redshift correlation can be related to the radial coordinate separation by means of the redshift equation (\ref{eq:null})
\begin{equation}
 \Delta z_{BAO} = \int 
\frac{dz}{dr} dr \approx \frac{(1+z) \dot A'(r(z),t(z))}{\sqrt{1-k(r)}} \Delta r_{BAO}\,,
\end{equation}
where in the second equality the integrand has been assumed to be constant. Similarly, the coordinate characteristic scale $\Delta r$ is given in terms of the physical scale through an integral
\begin{equation}
 \Delta r_{BAO} \approx \frac{\sqrt{1-k(r)}}{A'(r(z),t(z))} l_{BAO}^R(z)\,.
\end{equation}
Both equations relate the physical correlation with the redshift correlation
\begin{equation}\label{delz}
\Delta z_{BAO} = (1+z) H_R(z)  l^R_{BAO}(z)
\end{equation}
Constructing the geometric mean (\ref{deltaz}) using (\ref{dtheta},\ref{delz}) is straightforward:
\begin{equation}\label{baoeq1}
d_z^{LTB} = \left( \frac{H_R}{z} (1+z) \frac{1}{D_A(z)^2} \right)^{1/3} \xi(z)\, l(r_\infty,t_0)\,,
\end{equation}
where the scale conversion arising from (\ref{lT},\ref{lL}) has been introduced in the factor 
$\xi(z) \equiv (\xi_R(z)\xi_T^2(z))^{1/3}$, given by
\begin{eqnarray}
\xi(z) &=& \left( \frac{A'(r(z),t(z))}{A'(r(z),t_e)} \frac{A'(r_\infty,t_e)}{A'(r_\infty,t_0)} \right)^{1/3}
\left( \frac{A(r(z),t(z))}{A(r(z),t_e)} \frac{A(r_\infty,t_e)}{A(r_\infty,t)} \right)^{2/3}  \,,
\end{eqnarray}
using a suitable early time $t_e= t(z\sim 100)$ to convert the scale as described in the previous section. Note that due to the FRW limit, the ratios of the factors computed at $r_\infty$ can be expressed as redshift factors $a(t_e)/a_0=(1+z_e)^{-1}$.

Equation (\ref{baoeq1}) can be easily related to the usual expression for $d_z$ 
\begin{equation}\label{baoeq2}
 d_z^{\rm LTB} = (1+z)\xi(z) \frac{l(r_\infty,t_0)}{D_V(z)}
= (1+z)\xi(z) \, d_z^{\rm FRW}\,.
\end{equation}
in terms of the usual volume distance 
\begin{equation}\label{DV}
 D_V(z)=\left((1+z)^2 D_A(z)^2 \frac{z}{ H_R(z) }\right)^{1/3}\,.
\end{equation}
Relation (\ref{baoeq2}) absorbs the effects of the inhomogeneous rescaling in the BAO observations, which pick up a factor $(1+z)\xi(z)$ with respect to the FRW case. The difference between the two rescaling factors accounts for the anisotropy between the transverse and longitudinal BAO scales, while their redshift dependence is a consequence of the inhomogeneity. Both effects are shown in Figure \ref{baorescaling}.
Note also that there will be an additional difference because of the modified relations between the angular diameter distance (related to the transverse expansion) and the longitudinal expansion rate entering the geometric mean distance (\ref{DV}).

\subsubsection{The Alcock-Paczynski effect in LTB models} \label{ap-test}

The Alcock-Paczynski (AP) effect \cite{Alcock:1979mp} is the geometric distortion of spherical objects due to cosmological expansion, since distances in the radial direction away from an observer are determined in redshift space, while transverse distances are seen as angular separations in the sky. This motivates the definition of the dimensionless {\it distortion factor}:
\begin{equation}\label{apeq}
 f^{\rm AP}_{_{\rm FRW}}(z)\equiv \frac{\Delta z}{\Delta \theta} = D_A(z) H_R(z) (1+z) \,.
\end{equation}
In a homogeneous universe, the above relation can be tested against spherical (or spherically distributed) objects for which $\Delta z$ and $\Delta \theta$ are measured.\footnote{In practice, the AP test is difficult to perform due to dynamical effects such as the redshift space distortions caused by cosmic structures, which induce peculiar velocities that affect the redshift in a systematic way.} This technique has been used to constrain cosmological models \cite{Blake:2011ep,Kazin:2011xt,Xu:2012fw}.

In LTB models with space-independent Big Bang time, initially spherical distributions are intrinsically distorted due to the local shear, as discussed in Section \ref{freefallbao}. Using the angular and redshift projection of physical distances for the inhomogeneous models given by equations (\ref{dtheta}, \ref{delz}), the analogue of the AP relation is modified by the ratio of the radial and transverse rescaling factors
\begin{equation}\label{apeq2}
 f^{\rm AP}_{_{\rm LTB}}(z) = \frac{\xi_R(z)}{\xi_T(z)} D_A(z) H_R(z) (1+z) = \frac{\xi_R(z)}{\xi_T(z)}  f^{\rm AP}_{_{\rm FRW}}(z)  \,.
\end{equation}
The values of these factors in both directions can be seen in Figure \ref{baorescaling}. As the universe expands faster in the transverse than in the radial direction, the distortion factor has a lower value than in FRW models, on top of the different relation between $D_A(z)$ and $H_R(z)$.

The distortion factor (\ref{apeq2}) is sensitive to cosmic shear (\ref{shear}) through the ratio of the transverse and radial rescaling (e.g. steeper profiles enhance the asymmetry). In the limit of zero background shear, the ratio of rescaling factors tends to one, and the only difference w.r.t. FRW comes from the different relation between the angular diameter distance and the radial Hubble rate.
Therefore, the information one obtains from the AP effect is {\it complementary} to the geometric mean distance given by Eq. (\ref{baoeq2}), which only depends on the expansion, i.e. the product of the rescaling in the three spatial directions, and is unable to tell apart $\xi_R$ from $\xi_T$. Therefore, the AP effect is not only able to distinguish FRW from LTB models, but could eventually allow to observationally discriminate between different LTB profiles.

\section{Observational Data}\label{section:observations}

The present analysis relies on the interplay between the cosmic distances obtained by type Ia supernovae and the distances \emph{and} rescaling constraints from the baryon acoustic oscillation scale. SNe can be regarded as a standard candle and BAO as an standard ruler, which suffers additional effects due to the inhomogeneity. The measurement of the local expansion rate and the CMB peaks are also considered, their effect being to provide a calibration for the standard candles and rulers, respectively. However, the main result is independent of this calibration.

\subsection{Type Ia Supernovae}

The dimming of distant supernovae constitutes a solid probe of void models in the interval $0.01\lesssim z \lesssim 1.5$, as the luminosity distance depends on all the parameters of the model in a nontrivial way. The difference in magnitude between each observed supernovae at redshift $z_i$ and the theoretical expectation given by the luminosity distance $D_L(z)$ is
\begin{equation}\label{magDL}
\mu^{\rm th}(z_i)-\mu^{\rm obs}_i = 5\log _{10} \left(\frac{D_L(z_i)}{1\mbox{Mpc}}\right) + 25 -\mu_0 - \mu_i^{\rm obs} \equiv  \Delta\mu_i -\mu_0 \,,
\end{equation}
where the last equality defines the value of $\Delta\mu_i$ which is used for the observational constraints. The quantity $\mu_0$ depends on the intrinsic luminosity of the supernova explosions and will be allowed to take arbitrary values. Determinations of $\mu_0$ will be used to constraint the local expansion rate (see below).

In addition to the intrinsic luminosity, two other unknown quantities are necessary to calibrate the supernovae measurements and obtain a standard candle. These are the stretch (the duration of the supernovae explosion) and the color (to account for dust extinction), which are assumed to be universal and introduce linear corrections on all the SNe
$\mu^{\rm obs}_i = \mu_{B,i} - \mu_0 + \alpha (s_i -1) - \beta c_i $.
These factors are calibrated by assuming an FRW-$\Lambda$CDM model, and should in principle be allowed to vary if the cosmology changes \cite{Clifton:2008hv,Nadathur:2010zm}. 
However, the result of the analysis should not vary significantly since the LTB models we are considering usually give luminosity distance curves very similar to the standard model. The present analysis also includes the covariance matrix between the supernova data, which adds information about this calibration procedure by taking into account the covariance between supernovae with similar color and stretch. 

\begin{figure}
\begin{center}
 \includegraphics[width=0.49\columnwidth]{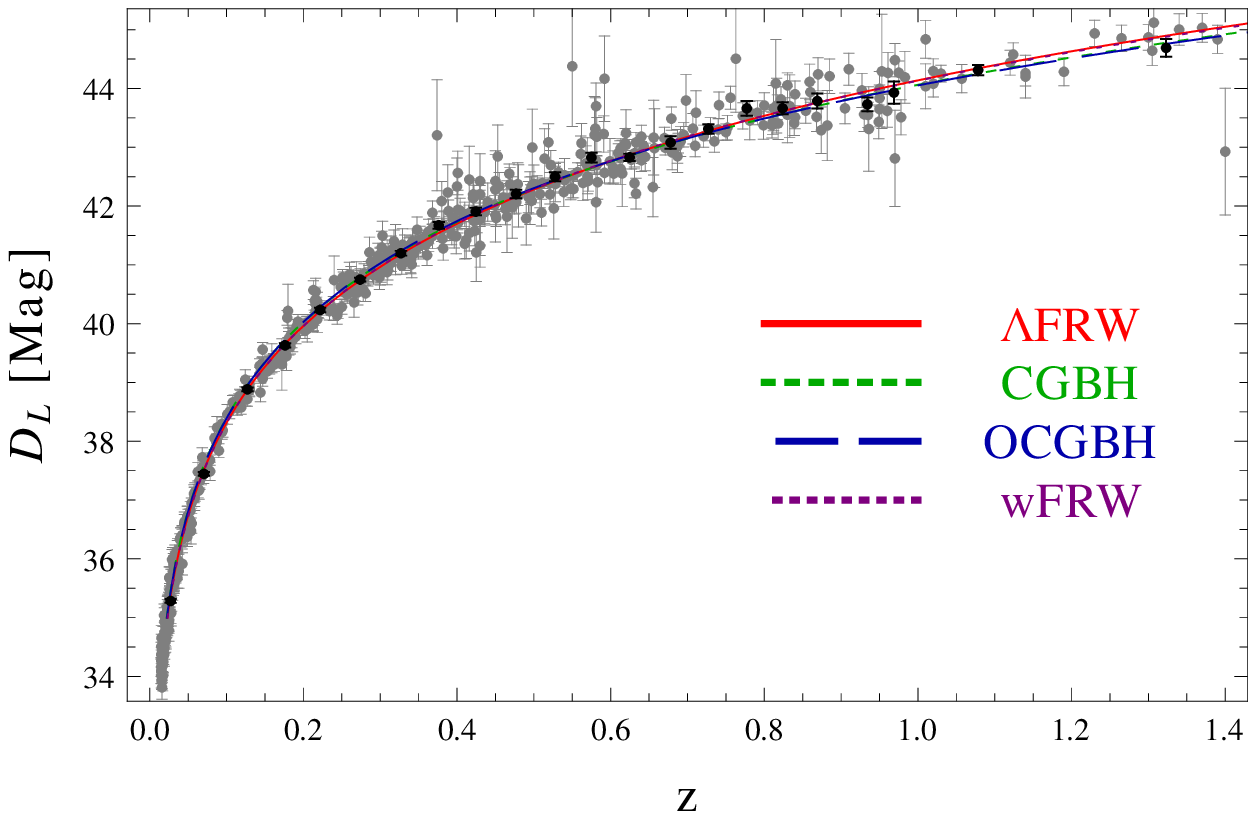}
\includegraphics[width=0.49\columnwidth]{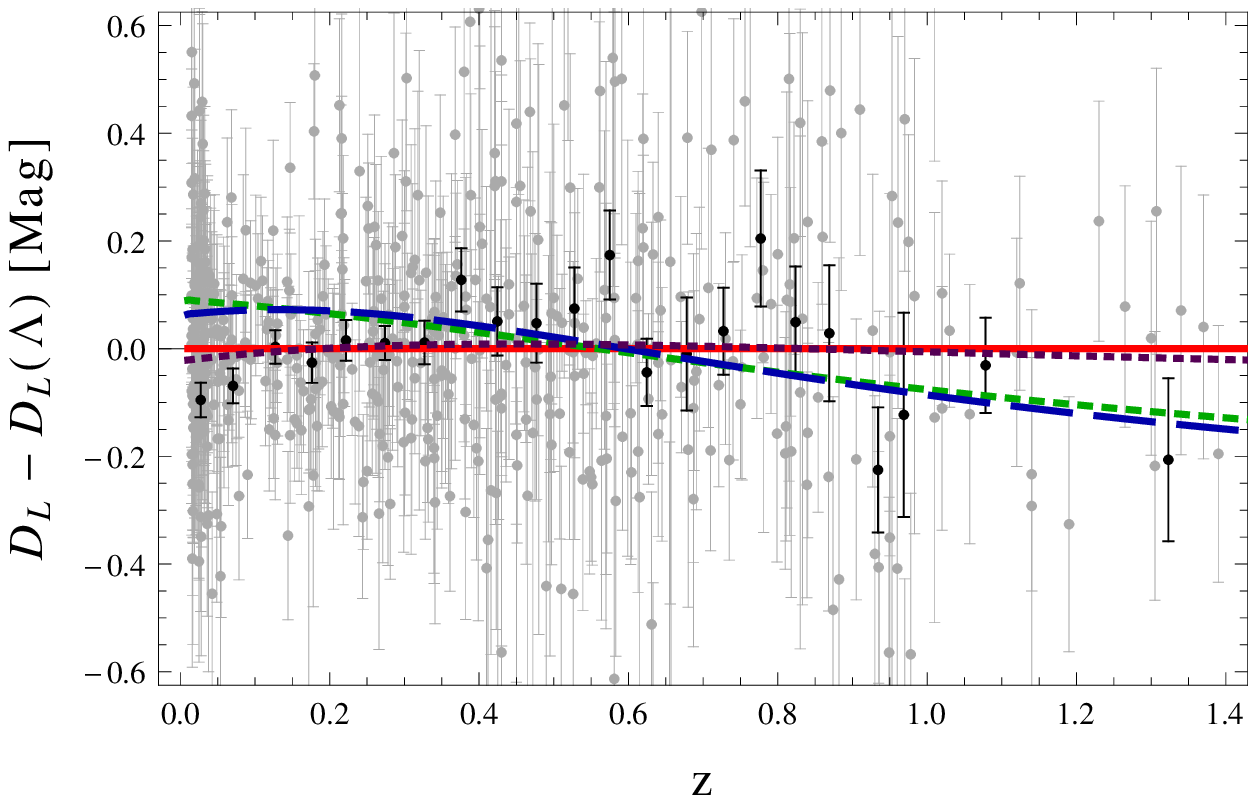}
\end{center}
\caption{Supernovae data and luminosity distance. The gray points correspond to the Union 2 compilation \cite{Amanullah:2010vv} used in the computation of the likelihood. The black points correspond to a binning of the same data using the covariance matrix, and are intended only for visual aid.
Color lines correspond to the minimum $\chi^2$ models described in Section \ref{results}, rescaled with the optimal value of $\mu_0$, as described in the text.
\label{sne}}
\end{figure}

The Union 2 supernovae compilation \cite{Amanullah:2010vv} consists in 557 SNe redshift-magnitude measurements after their lightcurves have been corrected for color and shape, shown in Figure \ref{sne}. The likelihood is computed using the covariance matrix including systematic errors $C_{ij}$
\begin{equation}\label{chisne1}
 -2\log L_{\rm SNe} = \chi^2_{\rm SNe} = \sum_{i,j} (\Delta\mu_i - \mu_0) C_{ij}^{-1} (\Delta\mu_j - \mu_0)\,.
\end{equation}
The above result depends on the actual value of the intrinsic luminosity $\mu_0$. Since it is unknown, the likelihood has to be maximized for each model with respect to $\mu_0$ for each model under consideration \cite{Kowalski:2008ez}.
Expanding the above expression and substituting back the value of $\mu_0$ such that $\partial (\chi^2) / \partial \mu_0 = 0$,
gives the optimal likelihood for each model
\begin{equation} \label{chi2sne} 
 \chi^2_{\rm SNe} = \sum_{i,j} \Delta\mu_i  C_{ij}^{-1} \Delta\mu_j 
- \frac{\left(\sum_{i,j} C_{ij}^{-1} \Delta\mu_j\right)^2}{\sum_{i,j} C_{ij}^{-1} }\,.
\end{equation}

\subsection{Local Expansion Rate}

Recasting the expression for the luminosity distance (\ref{magDL}) in units of $H_0$
\begin{equation}\label{magDL2}
\mu^{\rm th}(z_i) = 5\log _{10} \left(H_0 D_L(z_i)\right) + 25 -\mu_0 - 5\log_{10}(H_0 [\mbox{Mpc}^{-1}]) \,.
\end{equation}
it is possible to see that the intrinsic luminosity $\mu_0$ is degenerated with the Hubble constant for homogeneous models, for which cosmic distances only depend on it through a global $H_0^{-1}$ factor. Although inhomogeneous cosmologies introduce additional scales and allow for more involved dependences \cite{February:2009pv}, the determination of the local expansion rate requires the knowledge of the intrinsic supernovae luminosity.

A recent measurement of the local expansion rate using Ia type supernovae yields a value $H_0=73.8 \pm 2.4 \mbox{km}\mbox{s}^{-1}\mbox{Mpc}^{-1}$ \cite{Riess:2011yx}.
The supernovae intrinsic luminosity was measured using over 600 Cepheid stars from eight nearby galaxies in which type Ia supernovae have been observed. The Cepheids are calibrated comparing their luminosity to three different distance estimates: 1) the geometric distance to NGC 4258 as obtained from water masers orbiting its central black hole, 2) trigonometric paralaxes to Cepheid stars in the Milky Way and 3) relating the distance to the Large Magellanic Cloud obtained from eclipsing binaries. 
The local expansion rate is obtained by finding the best fit for a fiducial FRW model with $\Omega_M=0.3$, $\Omega_\Lambda=0.7$ to 253 low redshift type Ia supernovae ($z<0.1$) using the measured intrinsic luminosities. In particular, the quoted value of $H_0$ is the average of the values obtained from the three different calibrations.

The dependence of the expansion rate with redshift in a LTB cosmology is in general very different than in the $\Lambda$CDM case, even for low redshifts $z<0.1$. In order to reproduce the method used in \cite{Riess:2011yx} and provide a more fair comparison, we implement the constraints on the model using supernovae luminosities rather than the model parameter $H_{\rm in}$. The value and the error in the luminosity were obtained by comparing the fiducial model fixing $H_0=73.8$ and $73.8\pm2.4$ to the Union2 data in the range $z<0.1$ (195 SNe) and finding the value of $\mu_0^{\rm obs}$ that gives the best fit, using equation (\ref{bestmu}) below. The result is
\begin{equation}
 \mu_0^{\rm obs}=-0.120\pm 0.071\,.
\end{equation}
The ``predicted'' intrinsic luminosity that can be compared to the observation is the best fit $\mu_0$ found using the Union 2 data for the model under investigation
\begin{equation} \label{bestmu}
 \mu_0^{\rm bf} = \frac{\sum_{i,j} C_{ij}^{-1} \Delta\mu_j}{\sum_{i,j} C_{ij}^{-1} }\,,
\end{equation}
using the distance modulus and the inverse covariance matrix of the data (see previous section and equations (\ref{magDL},\ref{chisne1})). The associated likelihood is assumed to be Gaussian
\begin{equation}\label{chi2mu0}
\chi^2_{H_0} = \frac{(\mu_0^{\rm bf} - \mu_0^{\rm obs})^2}{\Delta{\mu_0}^2}\,.
\end{equation}

\subsection{Baryon Acoustic Scale} \label{BAS}

Although the use of BAO to constrain LTB models has raised some criticism \cite{February:2009pv,Marra:2011ct}, we will rely on our results from Section \ref{baoltb} showing that the baryonic features remain at constant coordinate positions to a good approximation and relating the transverse and radial BAO scales at different redshifts to the asymptitic values.

\begin{figure}
\begin{center}
\includegraphics[width=0.49\columnwidth]{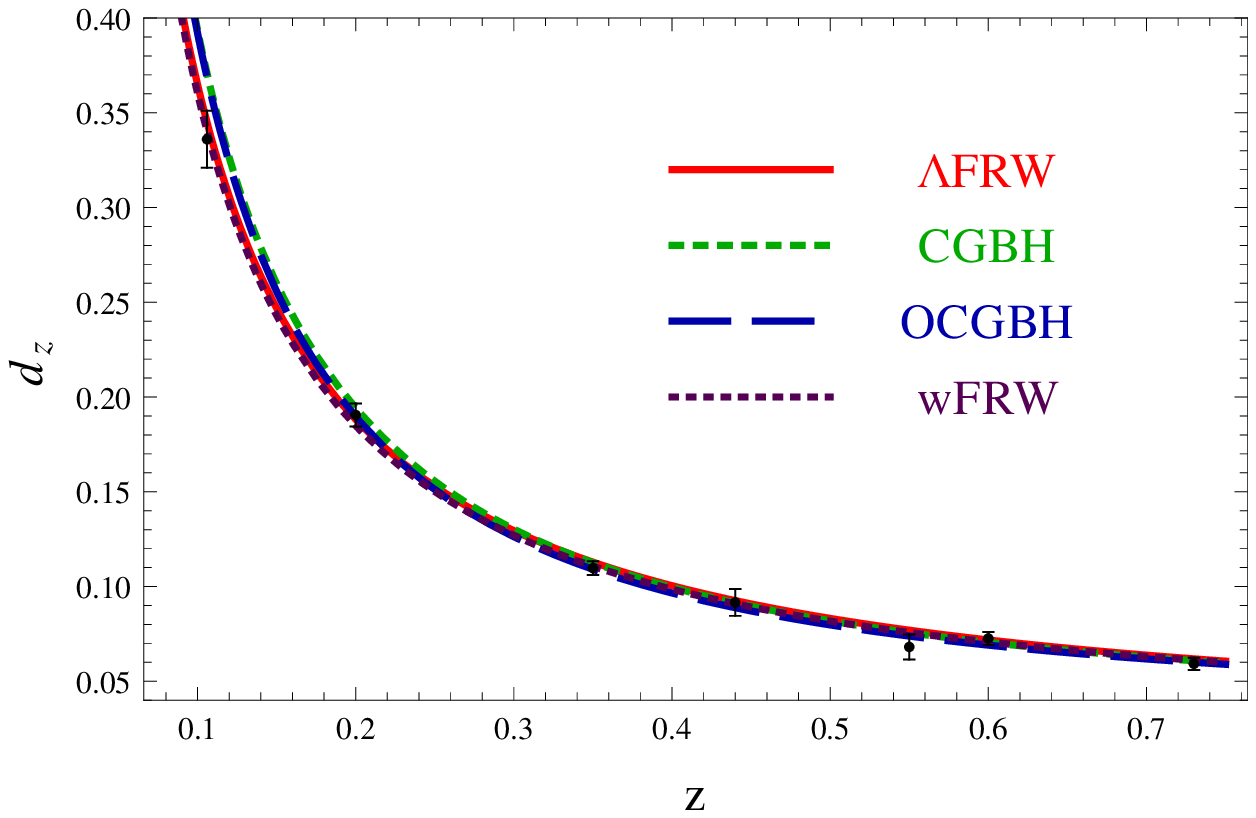} 
\includegraphics[width=0.49\columnwidth]{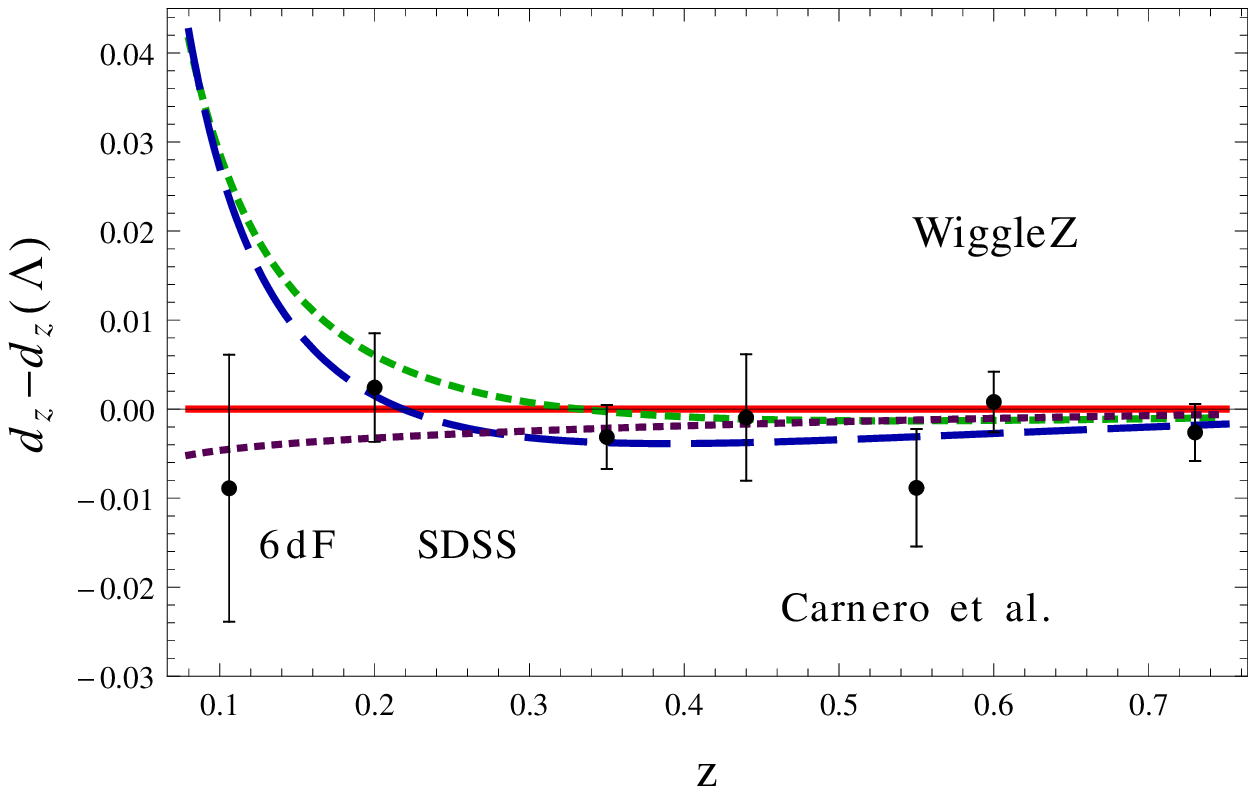} 
\end{center}

\caption{BAO data and $d_z$ for the best fit models described in Section \ref{results} and residuals with respect to $\Lambda$CDM. The point at $z=0.55$ has been converted from angular to volume distance by means of a fiducial model.
\label{bao}} 
\end{figure}

The WiggleZ collaboration \cite{Blake:2011en} has measured the baryon acoustic scale at three different redshifts, complementing previous data at lower redshift obtained by SDSS and 6DFGS \cite{Reid:2009xm,Percival:2009xn,Beutler:2011hx}.
Their measurements are given in terms of the variable $d_z$, which in our model is computed as (\ref{baoeq1}), or alternatively (\ref{baoeq2}), (\ref{DV}).\footnote{The $A(z)$ variable lacks an interpretation in the context of an LTB inhomogeneous model. However, we checked that using this method with values $\Omega_m|_{A(z)} = \Omega_m(z=0)|_{\rm LTB}\equiv \rho(r(0),t(0))/\rho(\infty,t(0))$ yielded consistent results.} An additional point involving the purely angular correlation \cite{Carnero:2011pu}, from SDSS DR7 catalog in the range $[0.5-0.6]$, has also been included at $z=0.55$, to be compared to the theoretical $\theta_{BAO}$ prediction (\ref{dtheta}).
All data points are summarized in Table 3 of \cite{Blake:2011en} and displayed in Figure \ref{bao} together with the best fit models.

\begin{table}
\begin{center}
 \begin{tabular}{| c | c | cc | ccc |@{\hspace{0.15cm}}| c | c |}
\multicolumn{1}{c}{} & \multicolumn{1}{c}{6dF} & \multicolumn{2}{c}{SDSS} 
 & \multicolumn{3}{c}{WiggleZ} & \multicolumn{2}{c}{Carnero \textit{et al.}} \\\hline
$z$ & 0.106 & 0.2 & 0.35 & 0.44 & 0.6 & 0.73 & $z$ & 0.55 \\\hline
$d_z$ & 0.336 & 0.1905 & 0.1097 & 0.0916 & 0.0726 & 0.0592 & $ \theta_{BAO}$ & $3.90^\circ $ \\\hline
$\Delta d_z$ & 0.015 & 0.0061 & 0.0036 & 0.0071 & 0.0034 & 0.0032 & $ \Delta\theta_{BAO}$ & $ 0.38^\circ $\\\hline
 \end{tabular} 
\end{center}
\caption{BAO data. The first six data points are volume averaged and correspond to Table 3 of \cite{Blake:2011en}. Their inverse covariance Matrix is given by (\ref{covbao}). The last point corresponds to an angular measurement given in \cite{Carnero:2011pu}. \label{baodata}}
\end{table}

The choice of data is convenient because it covers the redshift range $z\leq 0.8$ with a regular spacing and the correlations are known (see below). The point at $z=0.55$ was added to the ones summarized by the WiggleZ collaboration because it was obtained from the SDSS data in the interval $0.5<z<0.6$. Hence, it is independent of the measurements at $z=0.2,0.35$. Other available BAO scale determinations (e.g. references \cite{Sanchez:2008iw,Kazin:2009cj,Mehta:2012hh,Xu:2012fw,Anderson:2012sa}) would add points at intermediate redshifts with similar error bars and unknown covariances, and therefore we expect they will not increase the precision of the constraints. Determinations of the radial BAO scale \cite{Gaztanaga:2008xz,Gaztanaga:2008de} are of particular interest to constrain inhomogeneous models \cite{GarciaBellido:2008yq,Moss:2010jx} due to the distinct radial rescaling factor (\ref{lL}). Nonetheless, they were not included in the analysis due to the lack of knowledge about the correlations with other datapoints.

The likelihood is given by
\begin{equation} \label{chi2bao}
\chi^2_{\rm BAO}= \sum_{i,j} (d_i - d(z_i))C^{-1}_{ij}(d_j-d(z_j)) 
+ \frac{(\theta_{\rm BAO}(0.55)-\theta_{\rm BAO}^{0.55})^2}{\Delta\theta_{\rm BAO}^2}\,,
\end{equation}
where the indices $i,j$ are in growing order in $z$, as in Table \ref{baodata}.
For the first six points, $C_{ij}^{-1}$ was obtained from the covariance data in \cite{Blake:2011en} in terms of $d_z$:
\begin{equation}\label{covbao}
 C_{ij}^{-1}= \left(
\begin{array}{cccccc}
 4444 & 0. & 0. & 0. & 0. & 0. \\
 0. & 30318 & -17312 & 0. & 0. & 0. \\
 0. & -17312 & 87046 & 0. & 0. & 0. \\
 0. & 0. & 0. & 23857 & -22747 & 10586 \\
 0. & 0. & 0. & -22747 & 128729 & -59907 \\
 0. & 0. & 0. & 10586 & -59907 & 125536
\end{array}
\right)\,.
\end{equation}

\subsection{Cosmic Microwave Background} \label{cmb-section}

The cosmic microwave background radiation in LTB models has been actively investigated \cite{Yoo:2010qy,Yoo:2010ad,Clarkson:2010ej,Moss:2011ze,Zibin:2011ma}, as it constitutes the most solid piece of evidence for statistical isotropy and the most powerful tool in cosmological constraints. The obtention of precise constraints from the CMB is beyond the scope of this work and therefore only a relatively simple analysis based on the location of the first peaks will be employed, in order to give an idea of the effects of calibrating the standard rulers. This method yields weaker constraints than using the whole WMAP data and the spectra computed in linear perturbation theory, WMAP distance prior $R,l_a,z_*$ \cite{Komatsu:2010fb} (See Section \ref{results} and Figure \ref{contours-frw}) or other model independent determinations \cite{Vonlanthen:2010cd}. 

If our galaxy is located very close to the center of the void, the radiation coming from the CMB will be highly isotropic and therefore well described by the angular power spectrum $C_l$, with no direction dependence. As usual, it will display a characteristic pattern of peaks and troughs located at multipoles
\begin{equation}\label{peak-pos}
 l_m=\left(m-\phi_m\right)l_A\,,
\end{equation}
where integer values of $m$ labels the peaks, half integer values correspond to troughs, and $\phi_m$ are corrections that depends on the details of the cosmology \textit{before} the recombination epoch. The overall factor is fixed by the \emph{CMB acoustic scale}
\begin{equation}\label{acoustic-scale}
 l_A=\pi \frac{D_A(z_*)}{r_s(z_*)(1+z_*)^{-1}}\,,
\end{equation}
determined by the ratio between the observed angular diameter distance until recombination and the sound horizon at that epoch. Further information on the cosmological parameters can be obtained by considering the relative heights of the acoustic peaks compared to the first one $H_a=C_{l_a}/C_{l_1}$. 

The decoupling epoch occurs at an early time when the universe is very homogeneous and the primary anisotropies are produced on our past lightcone at a radius much larger than the size of the void $r(z\sim 1100)\gg R$.
In this case the pre-recombination physics is effectively the same as in a homogeneous universe, and we can assume that the relative peak positions $(m-\delta \phi_m)$ and heights $H_a$ are those of a FRW universe with the effective asymptotic values of the LTB model discussed in section Section \ref{earlybao}.\footnote{Variations in the effective CMB temperature have not been considered because for the profiles under consideration (compensated voids) there is no significant departure from $T_0=2.725K$ \cite{Marra:2010pg,Zibin:2008vk}.}
On top of modifying these asymptotic parameters, the only effects from the void will be to shift the peaks by varying the acoustic scale $l_A$ through the angular diameter distance $D_A(z_*)$.
Our analysis neglects secondary contributions such as the integrated Sachs Wolf effect on the lower multipoles or the action of gravitational lensing, which affects the relative heights of the peaks. Furthermore, we will assume no radiation contribution to the angular distance to recombination.\footnote{See references \cite{Clarkson:2010ej,Lasky:2010vn,Marra:2011zp} for discussions on radiation in the context of LTB models.}

\begin{figure}
\begin{center}
\includegraphics[width=0.49\columnwidth]{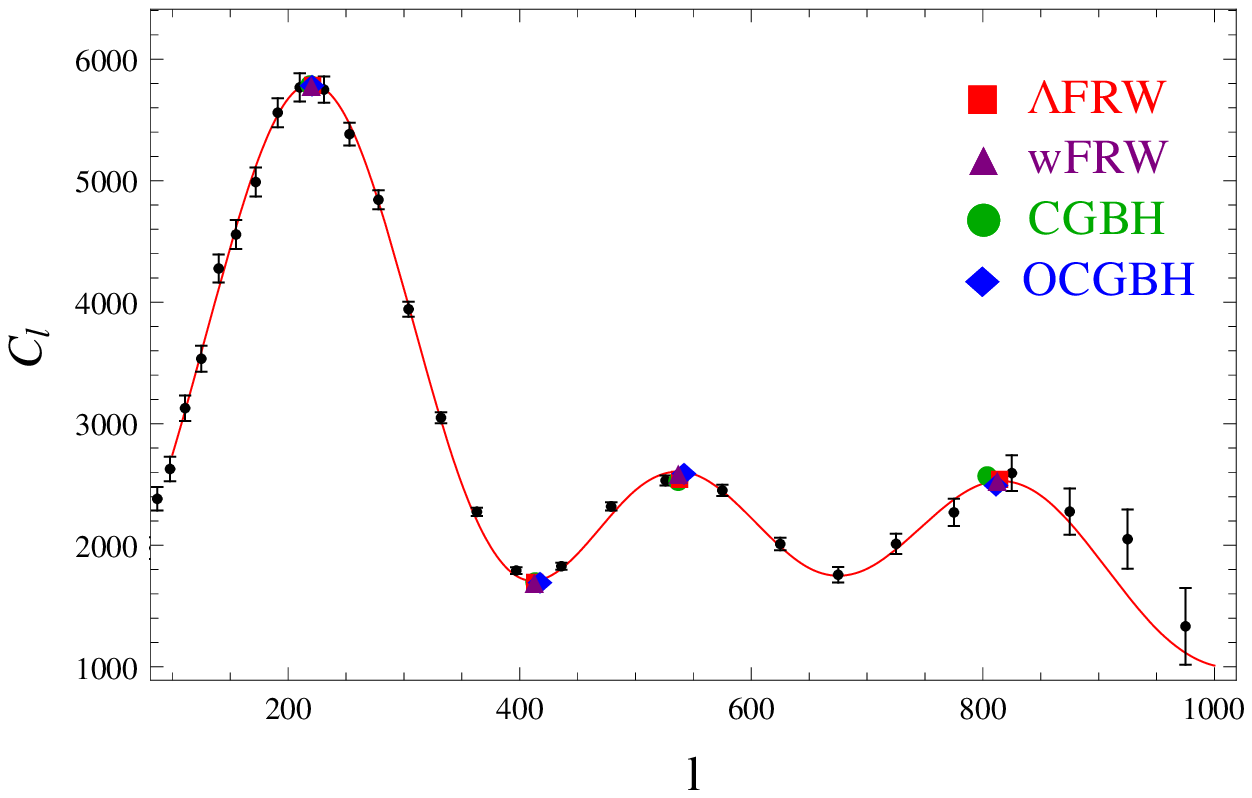} 
\includegraphics[width=0.49\columnwidth]{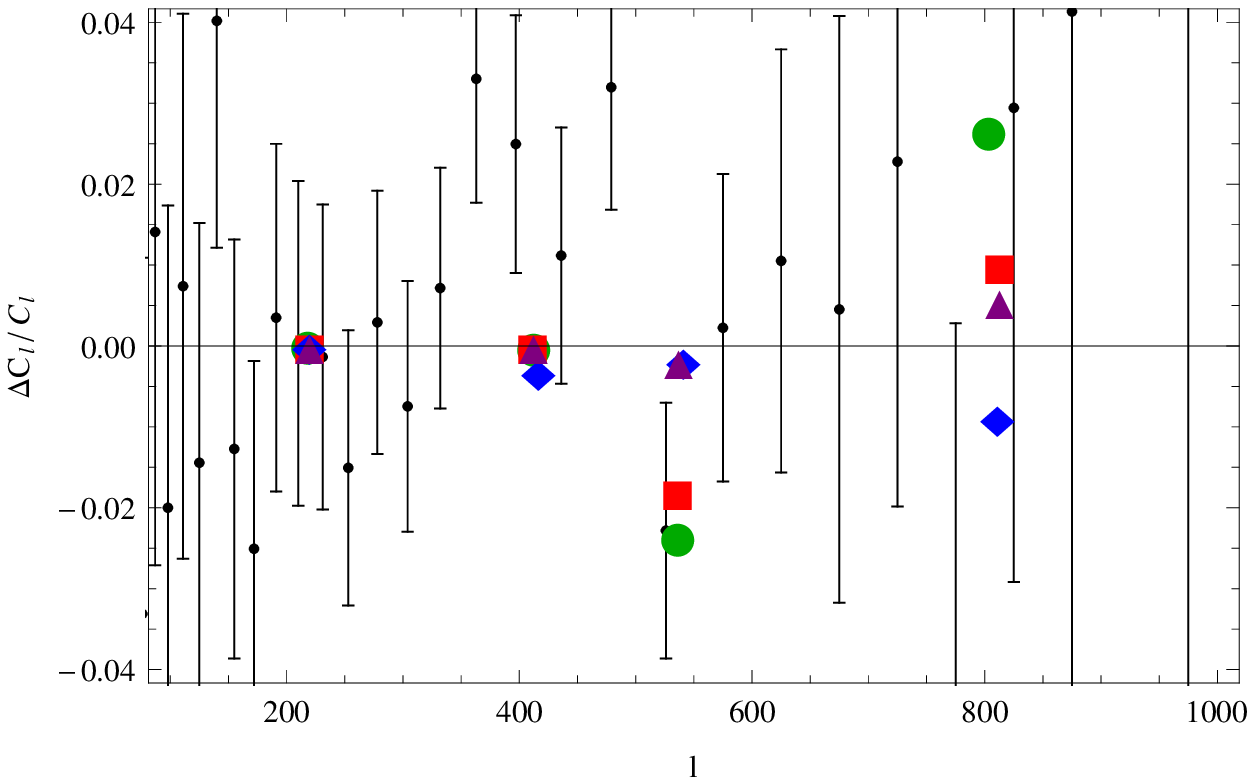} 
\end{center}
\caption{CMB spectrum. Black dots and error bars correspond to the binned WMAP7 data and the red line is the CMB only best fit for $\Lambda$CDM \cite{Larson:2010gs}. The color points are the reconstructed positions of the peaks using the method described in Section \ref{cmb-section} for the minimum $\chi^2$ models (Section \ref{results}). For visualization aid, the WMAP best fit height has been assumed for the first peak, which is equivalent to a normalization, and the first through due to the lack of a fitting formula (The residuals in $H_{3/2}$ are due to the propagation of $l_{3/2}$). Although the formulae do not exactly recover the values computed in linear perturbation theory, they fall within the assumed errors (1\% for $l_1$, 3\% for the rest). Note that the LTB models require values ($f_b\approx 0.7$, $n_s\approx 0.6$ that depart considerably from the standard model (Section \ref{results}). More precise constraints taken into account the full spectrum would considerably lower the quality of the fit. \label{cmb}} 
\end{figure}

In order to compare the theoretical predictions and the observations we will follow the approach described by Marra and P\"a\"akk\"onen \cite{Marra:2010pg}. The corrections to the peak locations $\phi_m$ and heights depend on the effective parameters through the ratio of matter-radiation density and recombination and the physical baryon density $\Omega_m h^2$, as well as the spectral index $n_s$ that characterizes the power spectrum of primordial perturbations. Note that relaxing the common assumption of a nearly scale invariant primordial spectrum considerably reduces the tension between CMB and the local expansion rate \cite{Nadathur:2010zm}.
Accurate fitting formulae in terms of these quantities are provided in reference \cite{Hu:1995en} for the recombination epoch $z_*$ and the sound horizon $r_s(z_*)$, reference \cite{Doran:2001yw} for $l_m$ with $m=1,\frac{3}{2},2,3$ and reference \cite{Hu:2000ti} for the relative height $H_a$ of the second and first peak $a=2,3$. Figure \ref{cmb} shows the location of the peaks reconstructed using this method.

The total likelihood is given by
\begin{equation}\label{chi2cmb}
 \chi^2_{\rm CMB}=\sum_{m\in \{1,\frac{3}{2},2,3 \}}\frac{\left(l_m^{\rm obs}-l_m^{\rm LTB}\right)^2}{2\sigma_{l_m}^2} 
+ \sum_{a\in \{2,3\}}\frac{\left( H_a^{\rm obs}- H_a^{\rm LTB}\right)^2}{2\sigma_{H_a}^{2}}\,,
\end{equation}
where the postions and heights of the peaks are those matching the WMAP 7 year best fit model. As in reference \cite{Marra:2010pg}, we have taken the errors to be of $1\%$ for the position of the first peak and $3\%$ for the remaining parameters.
It is important to note that this likelihood analysis is very simplified and its main aim is to provide an insight on how the information from the CMB helps to sharpen the BAO constraints by fixing the initial size of the standard ruler.

\section{Analysis and Results} \label{results}

In order to constrain the parameter space and address the viability of the different models, we run several Markov Chain Monte Carlo (MCMC) analysis using a modified version of the publicly available code CMBEasy \cite{Doran:2003sy}, which includes the integration of the coordinates over the lightcone and the computation of the cosmological observables in the LTB model described in sections \ref{baoltb} and \ref{section:observations}. CMBEasy's built in MCMC driver establishes the convergence of the chains through the test of Gelman and Rubin \cite{Gelman:1992zz}, which establishes the length of the burn-in sequence and freezes the step-size, which is a necessary condition for the convergence of the MCMC algorithm \cite{Doran:2003ua}. In addition, the chains were monitored manually to ensure a proper sampling of the parameter space.

Additionally to the CGBH and the OCGBH models described in section \ref{section_cgbh}, a $\Lambda$CDM model and a wCDM model with constant equation of state were studied using the same data.
Separate runs were performed for each of the displayed contours corresponding to the constraints of the separate sets (CMB, BAO, SNe) as well as the combined constraints H0+BAO+CMB+SNe. For the inhomogeneous models the additional combinations H0+SNe and BAO+CMB were considered, which combine the information of standard candles/rulers together with their calibrations (as opposed to the SNe/BAO-only). All the runs used flat priors on the model parameters, which are given in Table \ref{priors}.

\begin{table}[ht!]
\begin{center}
\vspace{0.5cm}
 {\sc FRW Models } \\ [0.2cm]
\begin{tabular}{|c c c c c c  |}
\hline
$H_0$ {\footnotesize [Mpc/km/s]} &$\Omega_{M}$ &$\Omega_{\Lambda}$ & $-w$ & $100f_b$ & $n_s$ \\ \hline
$30-90$ & $0.05-0.8$ & $0-1.2$ & $0-5$  & $1-25$ & $0.05-1.3$ \\ \hline
\end{tabular}
\vspace{1cm}

 {\sc GBH-LTB Models } \\ [0.2cm]
\begin{tabular}{|c c c c c c c |}
\hline
$H_{\rm in}$ {\footnotesize [Mpc/km/s]} & $\Omega_{\rm in}$ &$\Omega_{\rm out}$ & $R$ [Gpc] & $\Delta R$ [Gpc] & $100f_b$ & $n_s$ \\ \hline
$30-90$ &  $0.01-0.5$ & $0.1-1$ & $0-5$ & $0.5-5$ & $1-25$ & $0.05-1.3$ \\ \hline
\end{tabular}
\end{center}
\caption{Priors on the model parameters used in the MCMCs. In order to facilitate the comparison between the two LTB models, for the CGBH profile we have fixed the value of $H(r=0)\equiv H_{\rm in}$ at the center of the void instead of the more obscure parameter $H_0$. 
$\Omega_{\rm out}$ and $w$ are only varied in the OCGBH and wCDM models.
\label{priors}}
\end{table}

\begin{table}[ht!]
\vspace{0.5cm}
\begin{center}
\vspace{0.5cm}
 {\sc $\Lambda$CDM Model } \\[0.3cm]
 
\begin{tabular}{|r  c c c c c  |}
 \hline
 & $H_{0}$ {\footnotesize [Mpc/km/s]} & $\Omega_{M}$ & $\Omega_\Lambda$  &  $100f_b$ & $n_s$ \\[0.1cm] \cline{1-6}

{\footnotesize All - Min $\chi^2$} & 70.7 & 0.28 & 0.72 & 17 & 0.97   \\[0.1cm]

{\footnotesize Marginalized} & $70.3^{ +1.7}_{ -1.5}$ & $0.27{\pm 0.03}$ & $0.73\pm 0.05$ &
$17 \pm 0.04$ & $0.99^{+0.06}_{-0.09}$ 
\\[0.1cm]  \cline{1-6}
\hline
\end{tabular}

\vspace{1cm}

{\sc wCDM Model}  \\[0.3cm]
\begin{tabular}{| c c c c c c |}
\hline
$H_{0}$ {\footnotesize [Mpc/km/s]} & $\Omega_{M}$ & $\Omega_\Lambda$  &  $w$ & $100f_b$ & $n_s$ \\[0.1cm] \hline

 72.8 & 0.32  & 0.66  & $-1.26$ & 12 & 0.87  \\[0.1cm] 

 73.5 $\pm$ 2.3 & 0.33 $\pm$ 0.04 & 0.64 $\pm$ 0.06 
& $-1.26^{+0.17}_{-0.22}$ & $0.10^{+0.03}_{-0.02}$ & $0.82^{+0.08}_{-0.06}$ \\[0.1cm]
\hline 
\end{tabular}

\vspace{1cm}
 {\sc asymptotically flat Constrained GBH model (CGBH)}\\[0.3cm]
\begin{tabular}{|r  c c c c c c |}
\hline
& $H_{\rm in}$  & $\Omega_{\rm in}$ & $R$ {\footnotesize [Gpc] } & $dR$ {\footnotesize [Gpc]} & $100f_b$ & $n_s$ \\[0.1cm] \cline{1-7}
{\footnotesize All - Min $\chi^2$} 
& 66.4 & 0.21 & 0.02 & 2.78 & 7.7 & 0.74 \\[0.1cm]
{\footnotesize Marginalized} 
& $66.0{\pm 1.4}$ & $0.22 \pm 0.04 $ & $0.18^{+0.64}_{- 0.18}$ 
& $2.56 ^{+0.28}_{-0.24} $ & $7.7 \pm 0.4$ & $0.74 \pm 0.03$ \\[0.1cm] 
 \cline{1-7} 
\\[-0.4cm]
{\footnotesize BAO+CMB}  & 
$61.6{\pm 2.4}$ & $0.32^{+0.06}_{-0.04} $ & $3.92^{+0.48}_{-3.71}$ & $2.76 ^{+0.50}_{-0.88} $ & $7.8 \pm 0.8$ & $0.73 \pm 0.04$ \\[0.1cm] 
{\footnotesize SNe+H0}
 & $74.0{\pm 2.6}$ & $0.07\pm 0.04 $ & $1.95 ^{+1.22}_{-1.82}$ & $3.19 ^{+1.63}_{-1.66} $ & - & - \\[0.1cm] 
 \cline{1-7}
 \hline
\end{tabular}
\vspace{1cm}

{\sc asymptotically Open Constrained GBH model (OCGBH)} \\[0.3cm]

\begin{tabular}{| c c c c c c c |} 
\hline
$H_{\rm in}$ & $\Omega_{\rm in}$ & $\Omega_{\rm out}$ &  $R$ {\footnotesize [Gpc] } & $dR$ {\footnotesize [Gpc] } & $100f_b$ & $n_s$ \\[0.1cm] \hline

71.8 & 0.21  & 0.87  & 0.30  & 1.48 & 6.3 & 0.67 \\[0.1cm]
$71.1\pm 2.8$ & $0.22{ \pm 0.04}$ & $0.86\pm0.03$ 
&  $0.20^{+0.87}_{- 0.19}$ & $1.33 ^{+0.36}_{-0.32} $ 
& $6.2 \pm 0.5$ & $0.68 \pm 0.03$ \\[0.1cm]  \hline \\[-0.4cm]

$63.8^{+4.2}_{-2.8}$ & $0.35{ \pm 0.06}$ & $0.98^{+0.02}_{-0.11}$ 
&  $0.72^{+2.5}_{- 0.67}$ & $1.79 \pm 0.89 $ 
& $6.8 \pm 0.9$ & $0.69^{+0.05}_{-0.03}$ \\[0.1cm] 

$73.4^{+3.1}_{-2.1}$ & $0.06{\pm 0.04}$ & $0.89^{+0.09}_{-0.25}$ 
&  $0.80^{+1.66}_{- 0.74}$ & $1.63 ^{+2.04}_{-0.79} $ & - & - \\[0.1cm]  \hline

\end{tabular}

\end{center}

\vspace{1cm}

\caption{Parameters from the MCMC including H0+SNe+BAO+CMB discussed Section \ref{results}. The first lines correspond to the minimum $\chi^2$ models, while the second lines corresponds to the best fit model with one sigma errors after marginalizing over the remaining parameters. Note that since the LTB models do not give good fits to the data the errors are apparently very small (see Figures \ref{contours-frw}, \ref{contours-cgbh}, \ref{contours-ocgbh} and \ref{contours-ocgbh2}). For the sake of comparison, in the case of the LTB models the results from the separate fits using BAO+CMB and SNE+H0 have been added (third and fourth lines).
\label{table-parameters}}
\vspace{1.cm}
\end{table}

The results from the combined constraints can be seen in Table \ref{table-parameters}. Figures \ref{contours-frw}, \ref{contours-cgbh}, \ref{contours-ocgbh} and \ref{contours-ocgbh2} show the two-dimensional marginalized likelihood contours obtained from the individual and combined data sets. Our discussion starts by considering the homogeneous reference models. Then the results for the inhomogeneous CGBH and OCGBH profiles will be addressed, and the goodness of fit of the different models compared using different criteria.

\subsection{Homogeneous models}

\begin{figure}[ht!] 
\begin{center}
\textsc{$\Lambda$CDM Model}\\[0.8cm]
\includegraphics[width=0.45\columnwidth]{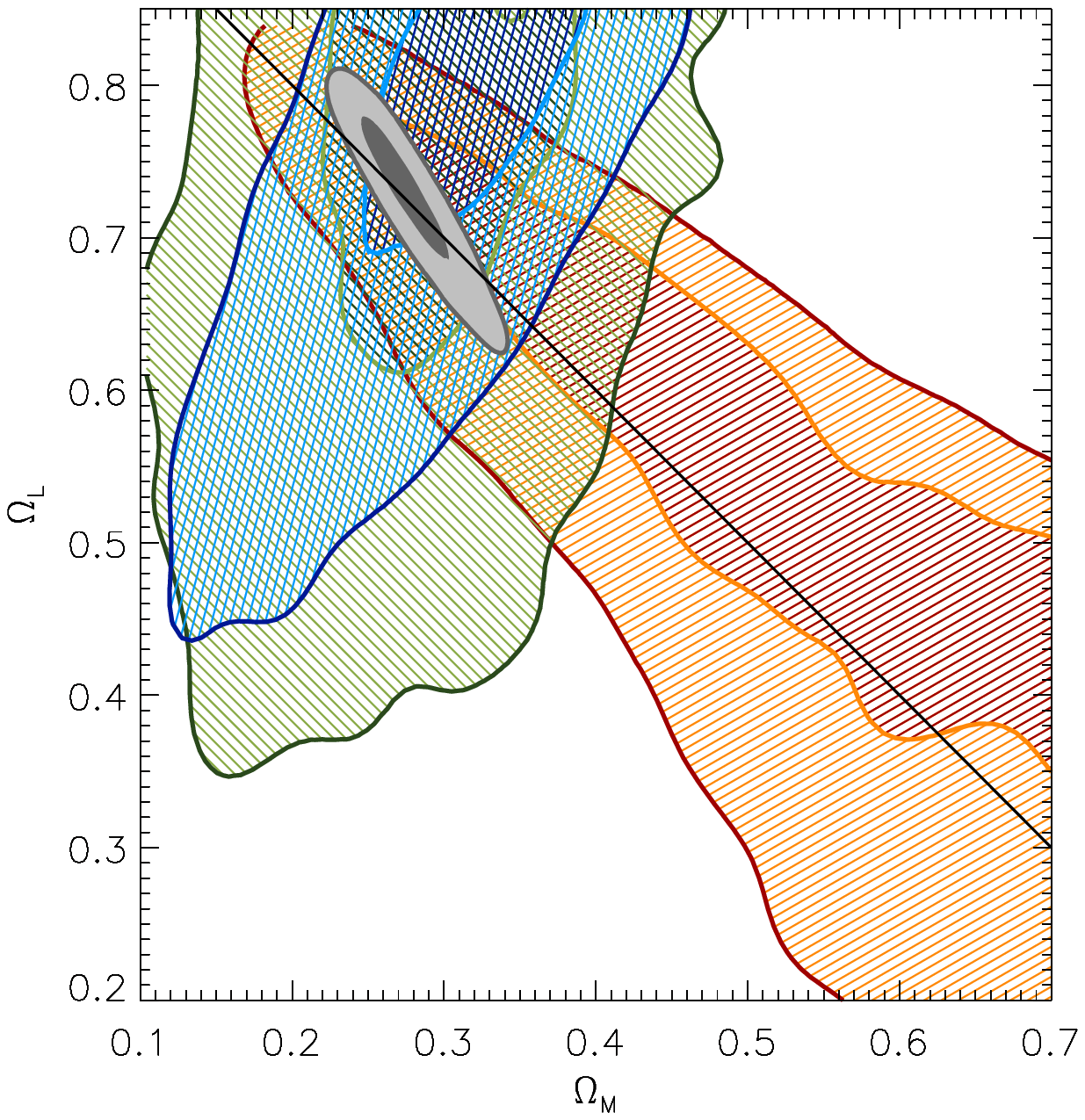} \hspace{0.48\columnwidth}
\\[1.4cm]
\textsc{wCDM Model}\\[0.8cm]
\includegraphics[width=0.45\columnwidth]{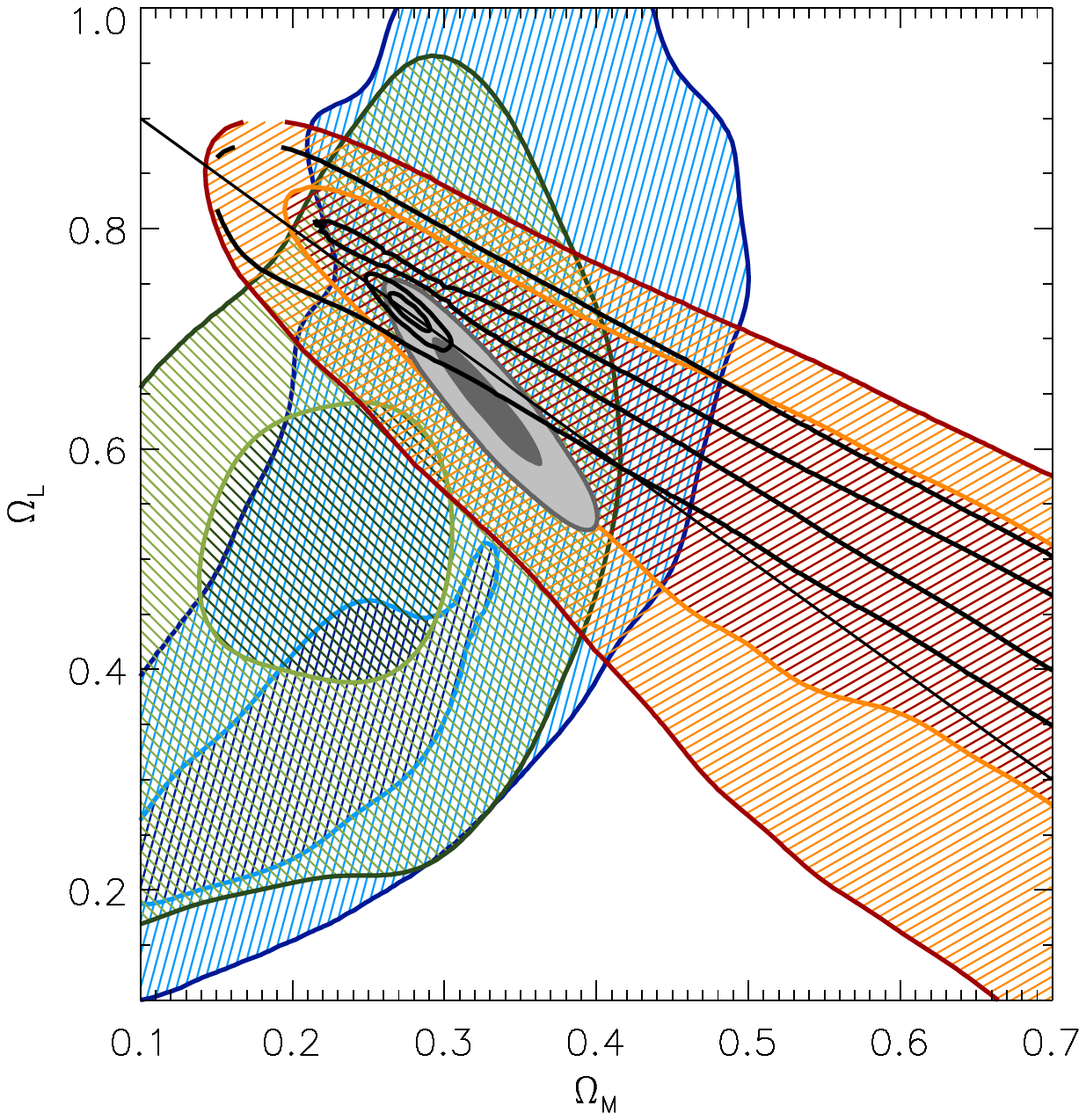} 
\includegraphics[width=0.45\columnwidth]{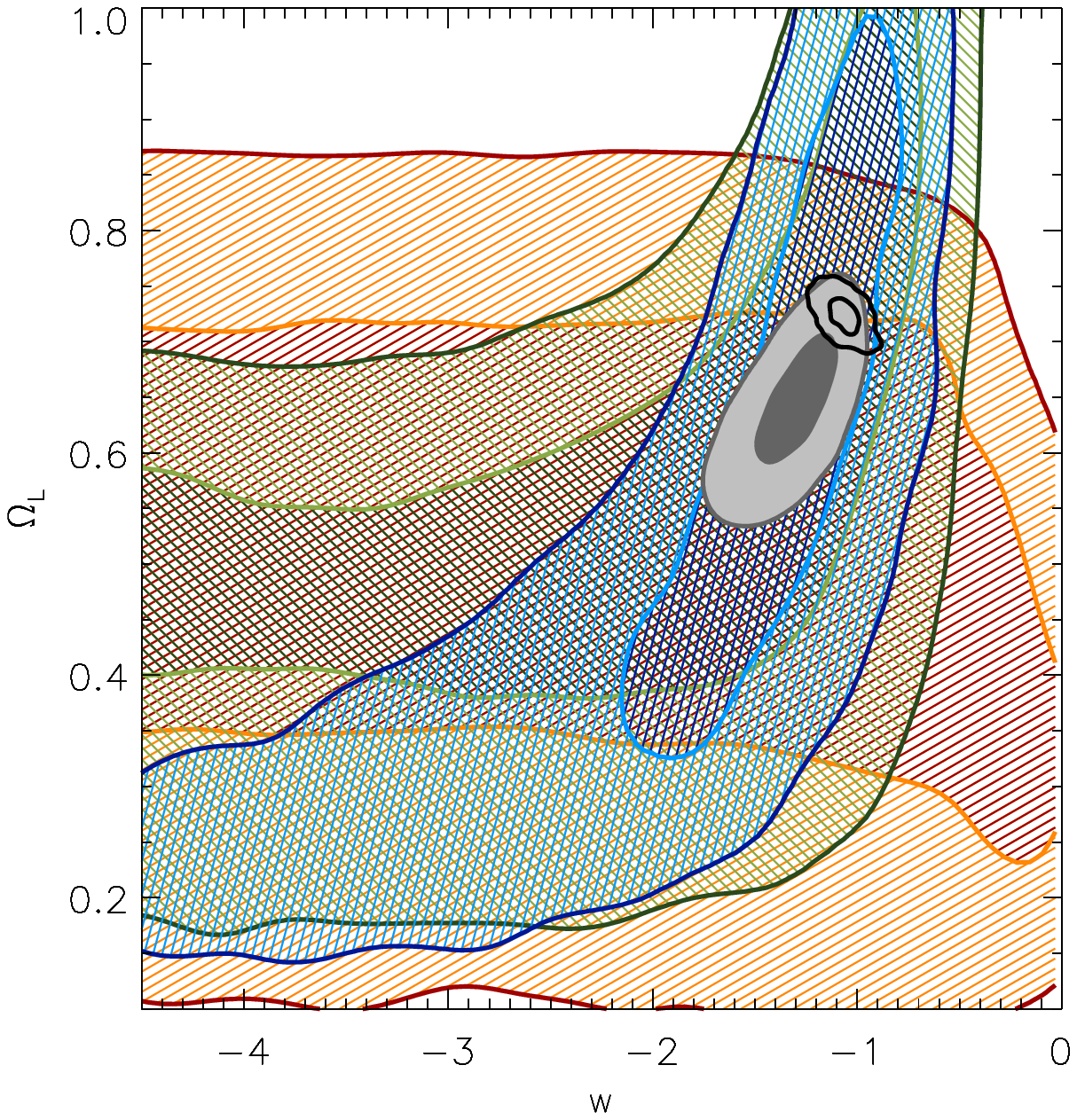} 
\end{center}
\caption{One and two sigma regions for the marginalized likelihood function corresponding to the $\Lambda$CDM and wCDM homogeneous model as obtained from BAO (green), CMB (Orange) and SNe (blue). Gray contours are the combined constraints H0+BAO+CMB+SNe.
Note that the CMB compatible regions are much broader than usual due to the simplification of the method.
Black unfilled lines in the wCDM plots correspond to using the WMAP distance prior $R,z_*,l_A$ \cite{Komatsu:2010fb} combined with H0+BAO+SNe or individually (shown only in the $\Omega_\Lambda-\Omega_m$ plane), which recover the standard results. \label{contours-frw}}
\end{figure}

For $\Lambda$CDM the recovered parameters are in good agreement with previous results. However, the region compatible with CMB data (first plot in Figure \ref{contours-frw}) is broader than usual around the flatness line. This lack of precision is caused by the partial use of the CMB data (i.e. only the peaks instead of the whole $C_l$ spectrum), together with the broad parameter priors allowed. When combined with other measurements, it also affects the recovered value of the curvature $\Omega_k = 0.003^{+ 0.015}_{- 0.025}$, which is still very close to flat but has larger error bars than usual. In a computation taking into account the full WMAP7 data, a deviation with respect to the measured values of the peak positions would also displace many of the intermediate points and cause a more dramatic decrease of the likelihood, leading to tighter bounds. 

The weakness of the CMB constraints is reflected again in the recovered values for the wCDM model (second and third plots in Figure  \ref{contours-frw}). In this case all the parameters except $H_0$, which is independently constrained by the nearby expansion rate, depart considerably from the standard ones. These allow for lower values of the baryon fraction and the spectral index, which in turn increase the matter fraction and decrease the dark energy content. A very dramatic consequence of the weakness of these constraints can be seen in the recovered value of the curvature $\Omega_k=0.04\pm 0.02$, two sigma away from flatness. The low value of $\Omega_\Lambda \approx 0.62$ is then compensated with an anomalously low equation of state $w\approx -1.34$.\footnote{The discrepancies disappear when the WMAP7 distance prior $R,z_*,l_A$ \cite{Komatsu:2010fb} are used instead of the peak positions (black, unfilled contours in Figure \ref{contours-frw}), recovering $\Omega_k\approx 0$ and $w\approx -1$. These quantities have been very accurately determined by the WMAP collaboration using the full CMB spectrum, and are able to break the degeneracies in the model (e.g. the baryon fraction). Although they are considerably more precise than the CMB peak information we used (described in Section \ref{cmb-section}), the WMAP distance prior can not be directly applied to inhomogeneous models (e.g. LTB models with decoupling redshifts $z_*\gtrsim 1110$ considerably higher than the standard value $z_*=1091.3\pm 0.9$ can yield a good fit \cite{Yoo:2010qy}).}

Note also how the BAO and SNe contours spann a similar region in both cases. This is a consequence of them being determined by measurements of standard rulers and candles with \emph{arbitrary} calibration, over a comparable redshift interval (due to the new BAO data provided by the WiggleZ collaboration, up to $z\sim 0.8$). Since in FRW both datasets depend only on the same distance-redshift relation and are consistent with each other, they yield basically the same information and the recovered regions overlap.

\subsection{Inhomogeneous models}

\begin{figure}[ht!]
\vspace{-0.3cm}
\begin{center}
\includegraphics[width=0.45\columnwidth]{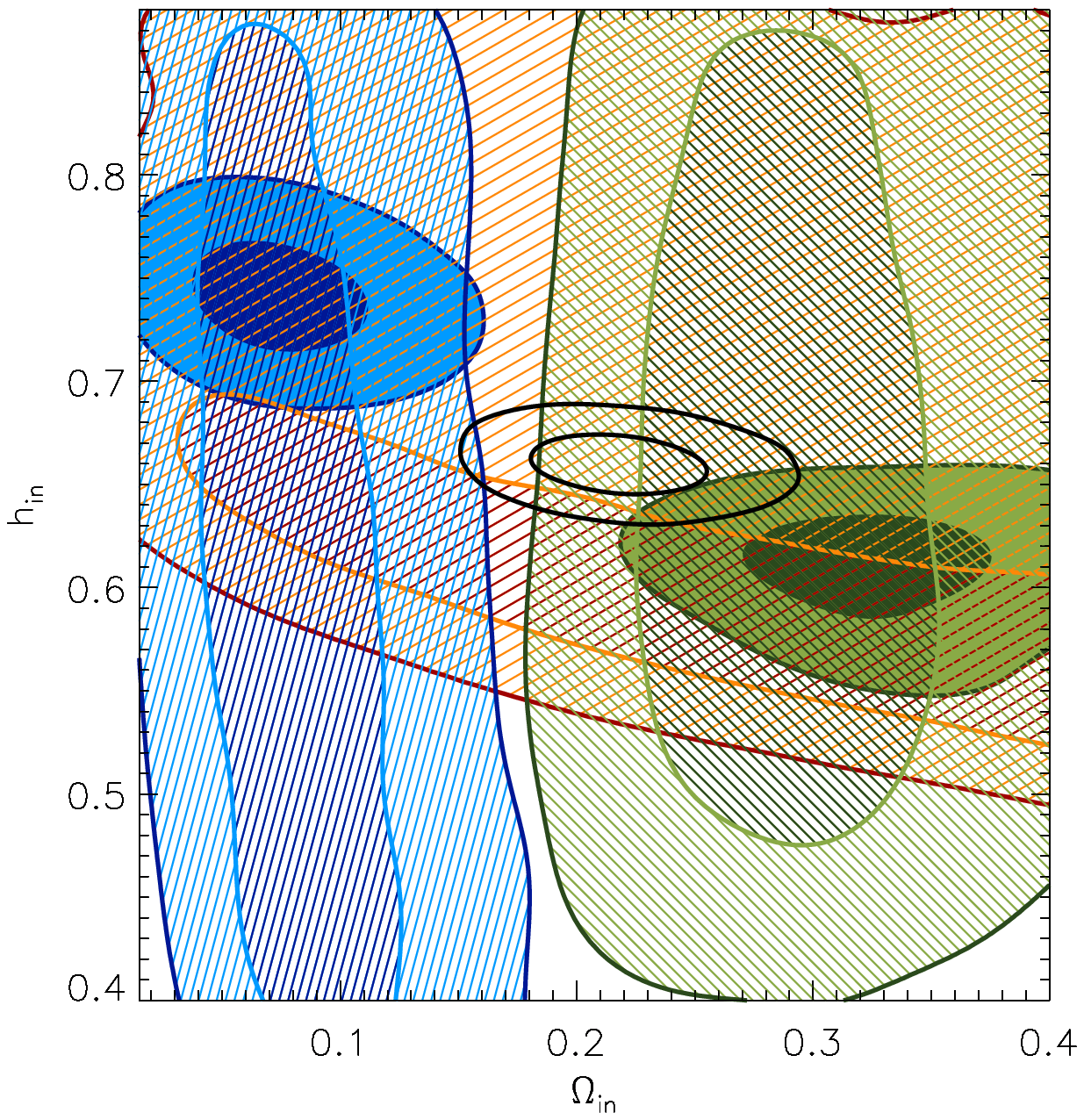}
\includegraphics[width=0.45\columnwidth]{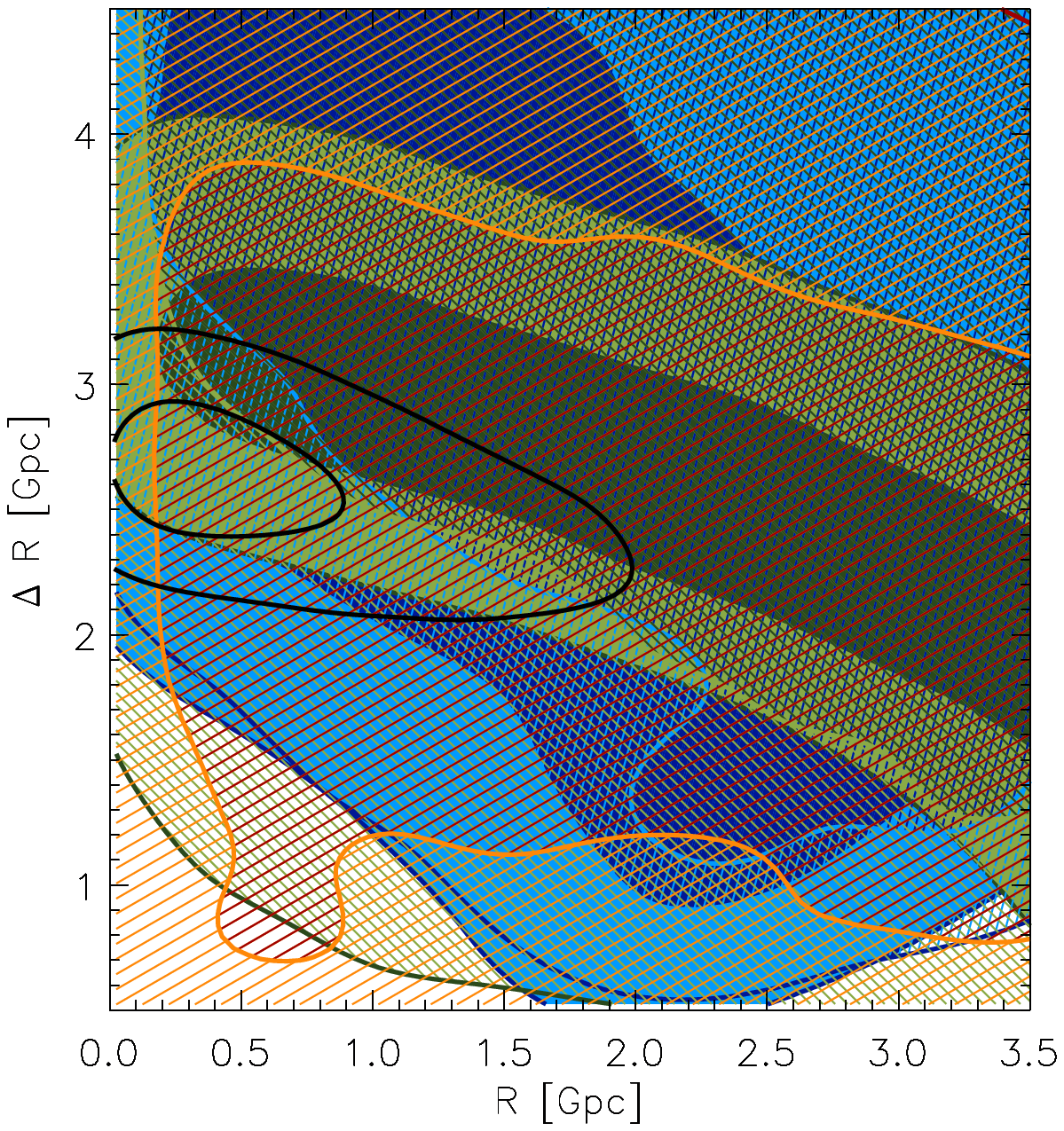}

\includegraphics[width=0.45\columnwidth]{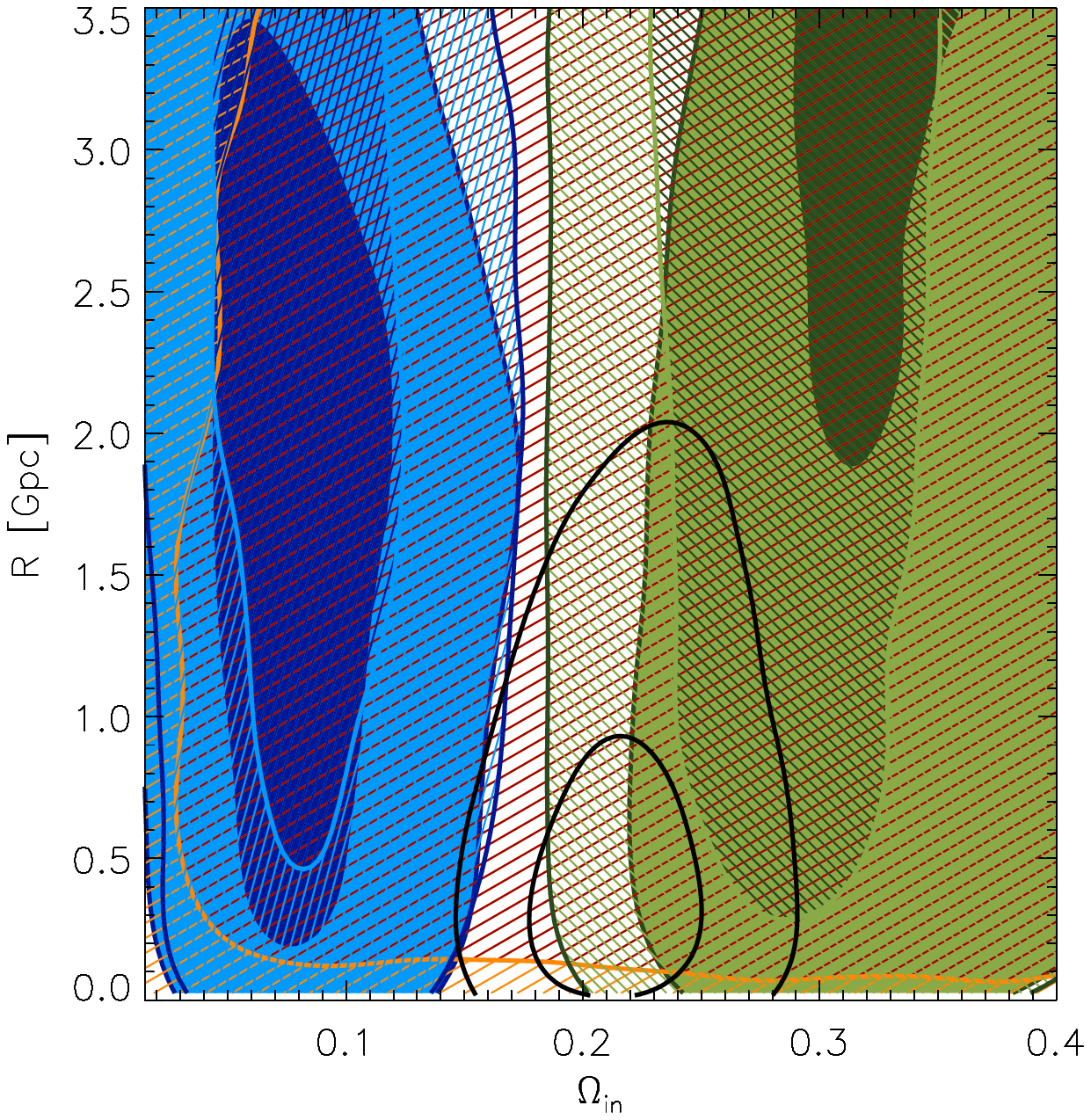}
\includegraphics[width=0.45\columnwidth]{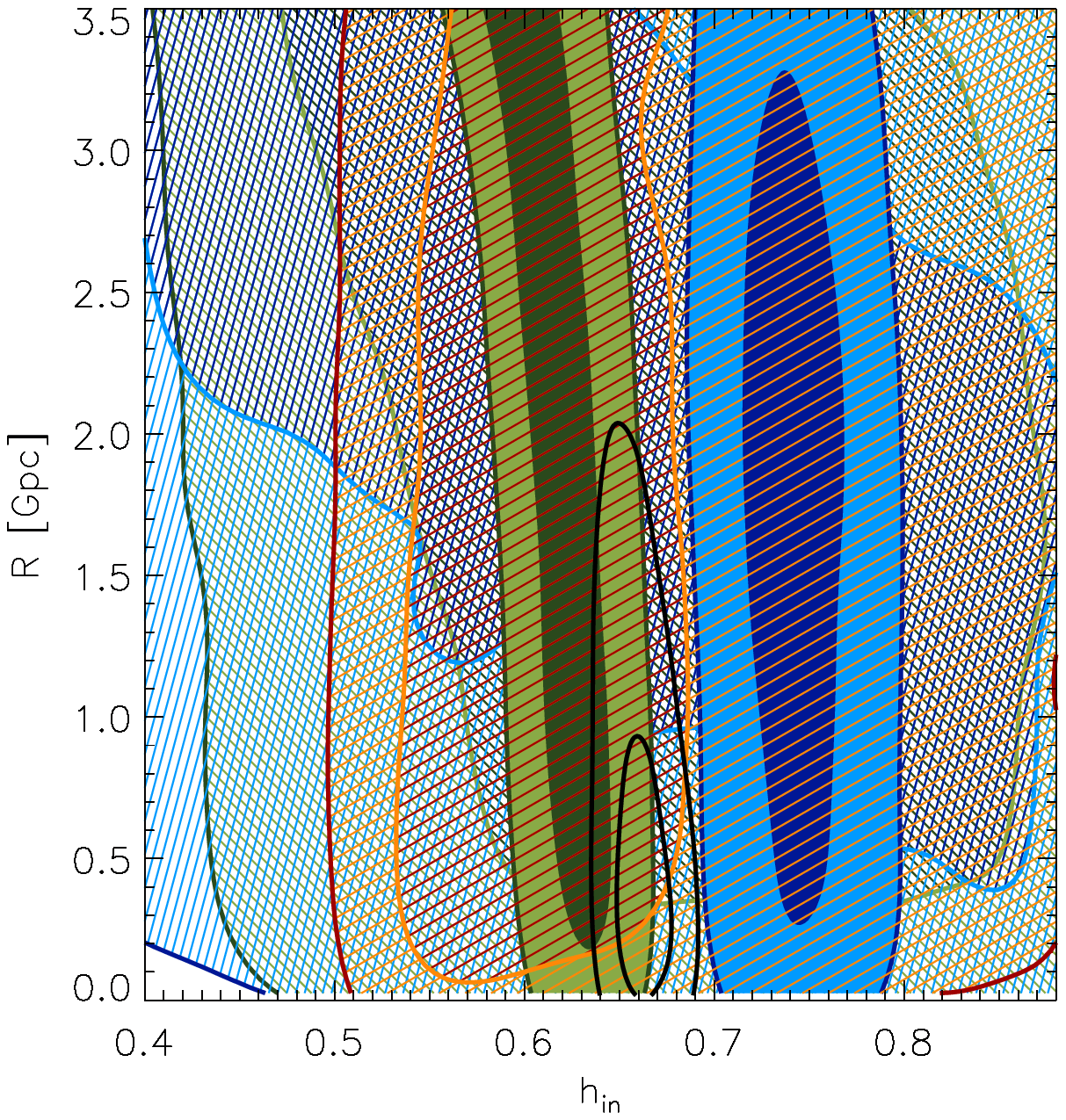}

\includegraphics[width=0.45\columnwidth]{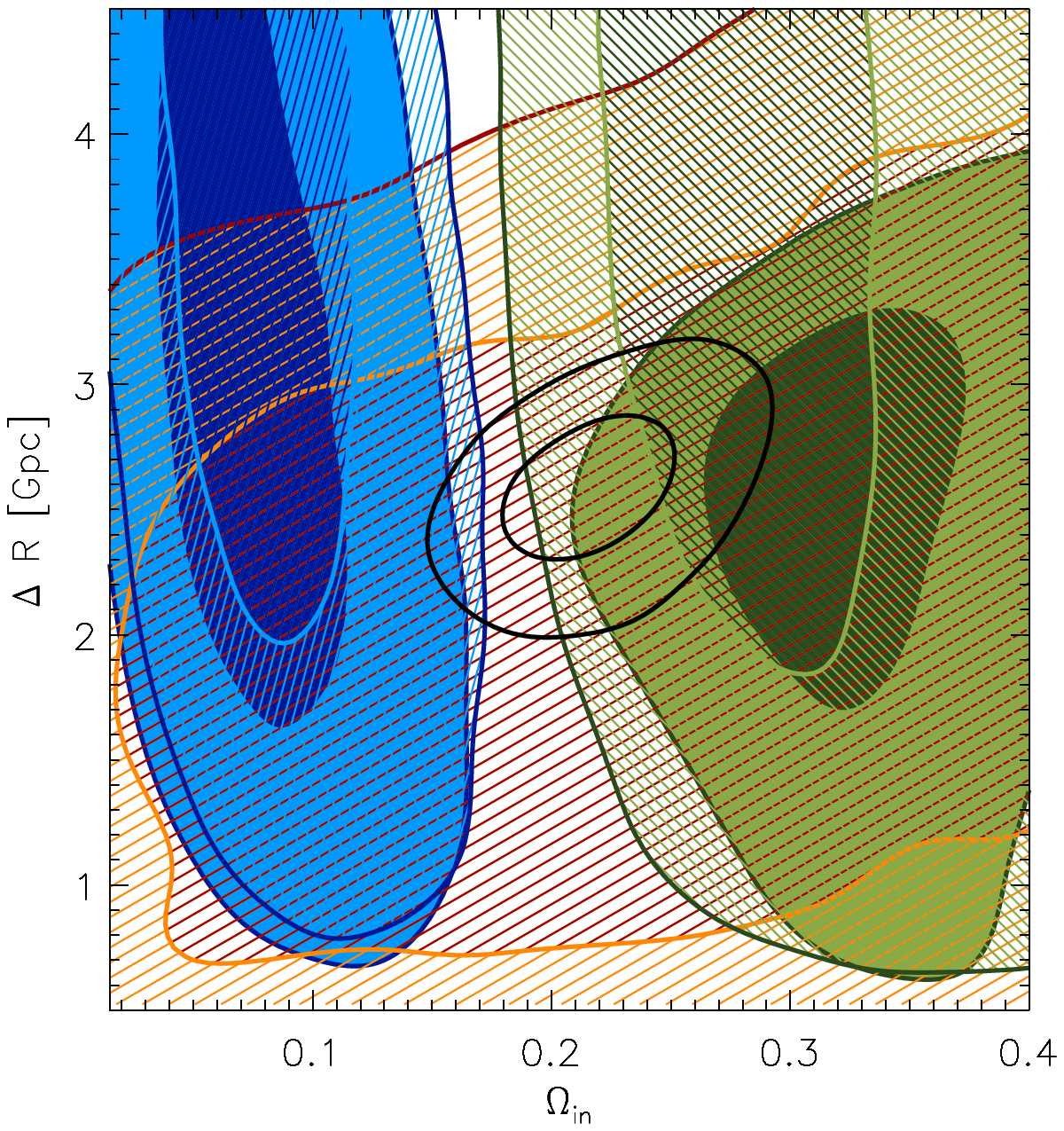}
\includegraphics[width=0.45\columnwidth]{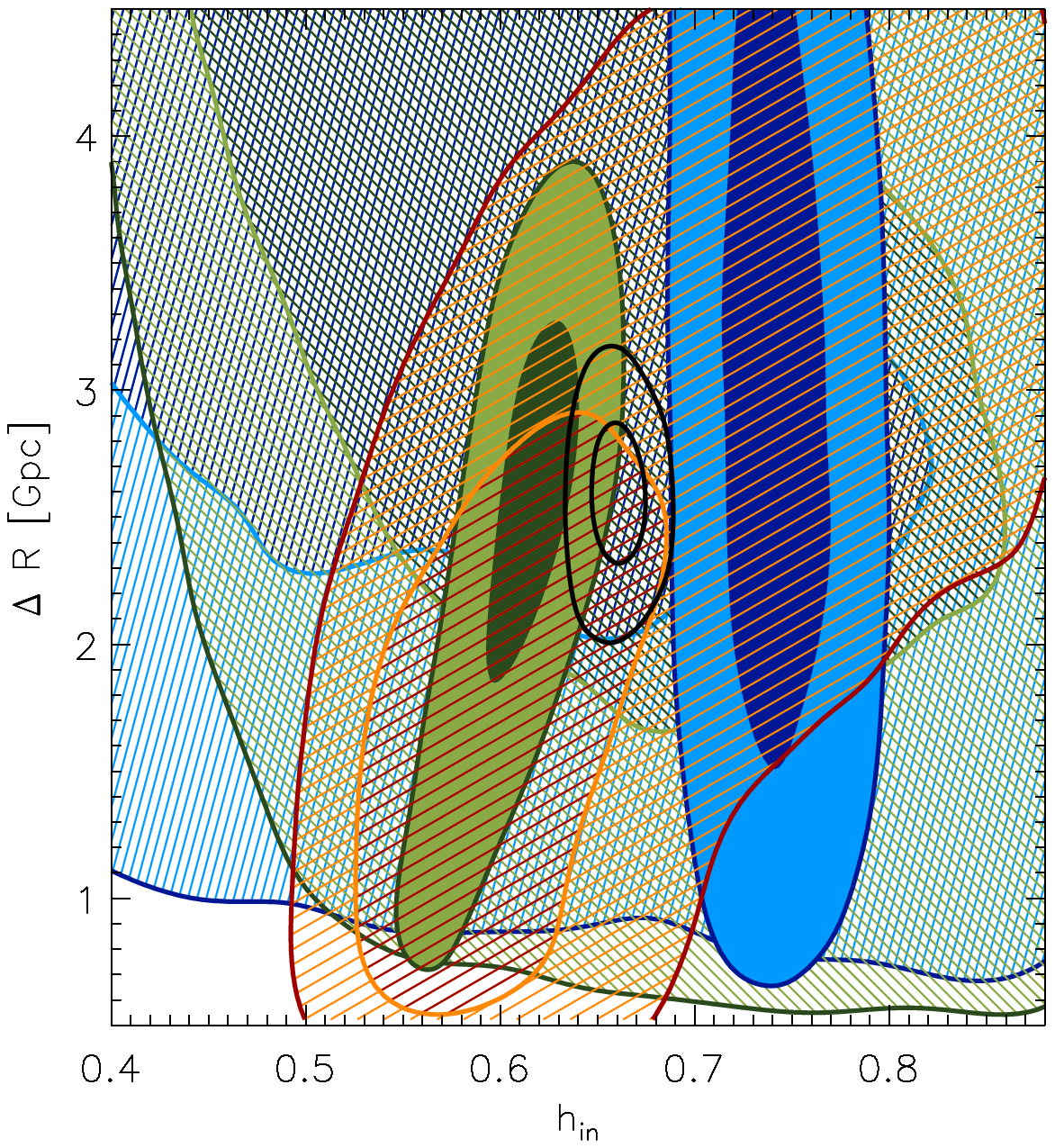}
\end{center}
\caption{CGBH model. One and two sigma regions for the marginalized likelihood function as obtained from BAO (green), CMB (Orange) and SNe (blue). The filled contours represent the combined constraints using SNe+H0 (blue) and BAO+CMB (green). The black lines correspond to the combined data BAO+CMB+SNe+H0.
\label{contours-cgbh}} 
\vspace{-1.5cm}
\end{figure}

Discrepancies between the different datasets are encountered for both models regarding the matter content and the expansion rate at the center of the void, as can be seen in Figures \ref{contours-cgbh}, \ref{contours-ocgbh}, \ref{contours-ocgbh2}. This becomes particularly clear for the $\Omega_{\rm in}-h_{\rm in}$ plane.

\begin{figure}[ht!]
\vspace{-0.3cm}
\begin{center}
\includegraphics[width=0.45\columnwidth]{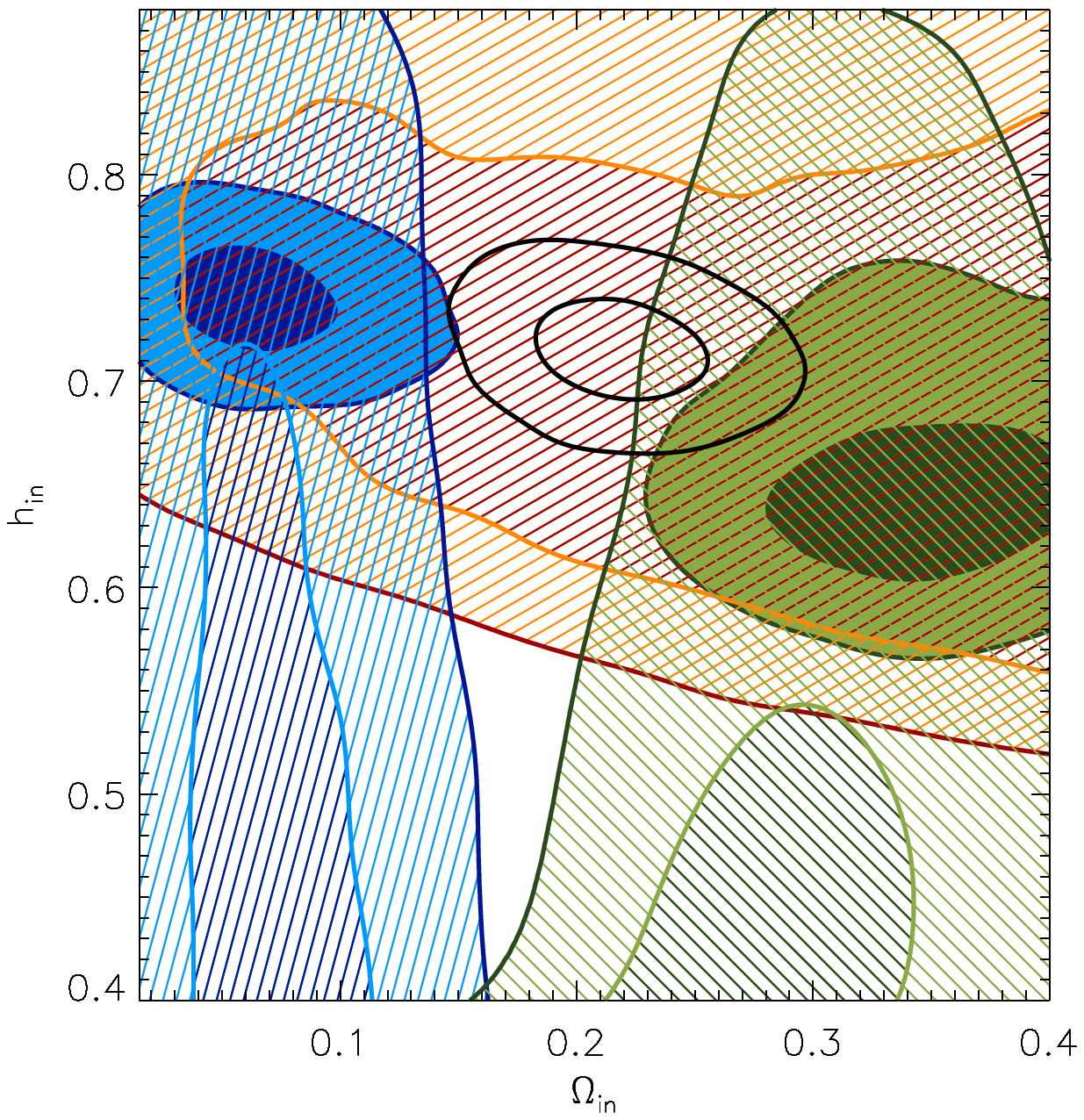} 
\includegraphics[width=0.45\columnwidth]{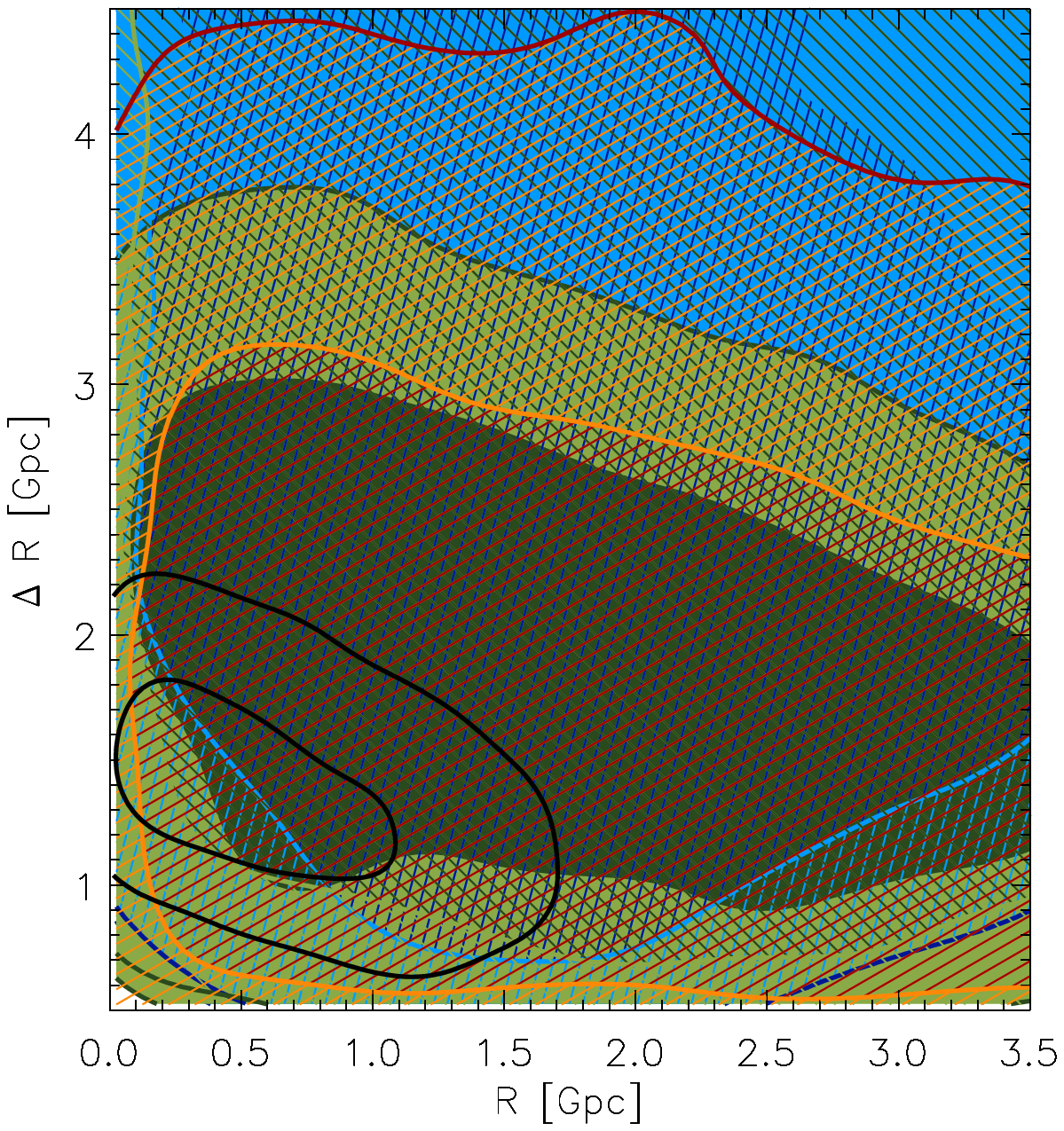}

\includegraphics[width=0.45\columnwidth]{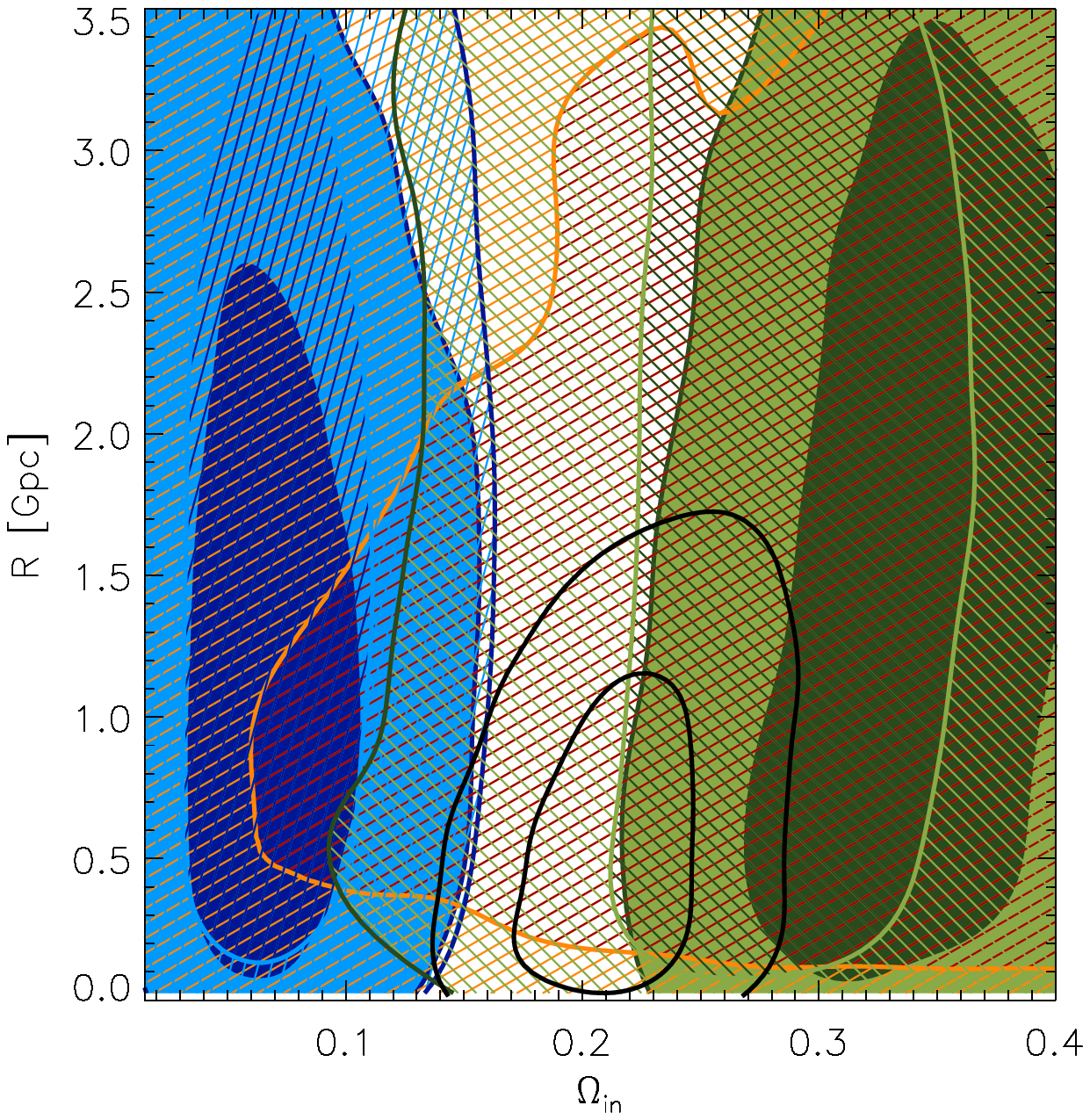}
\includegraphics[width=0.45\columnwidth]{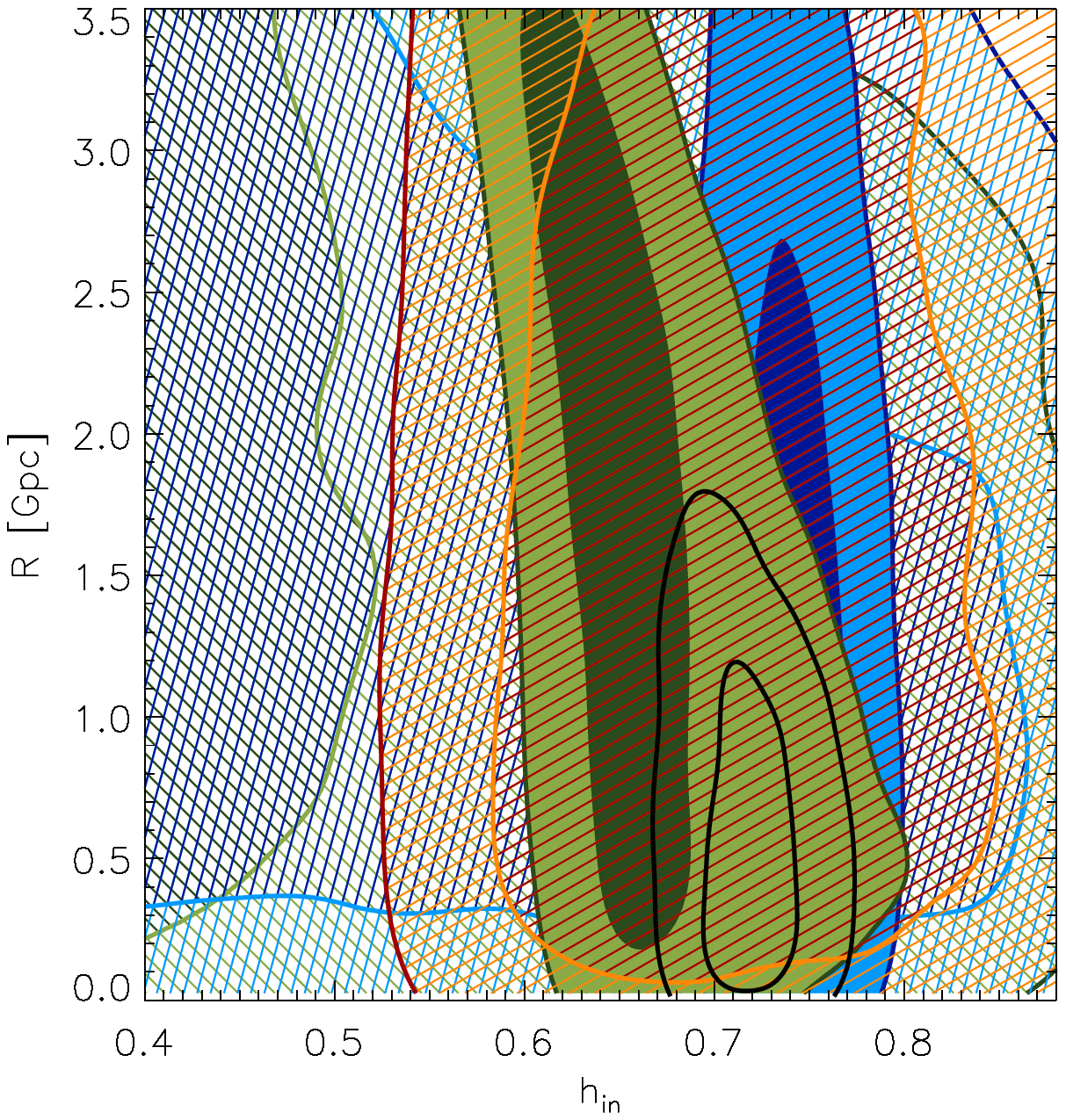} 

\includegraphics[width=0.45\columnwidth]{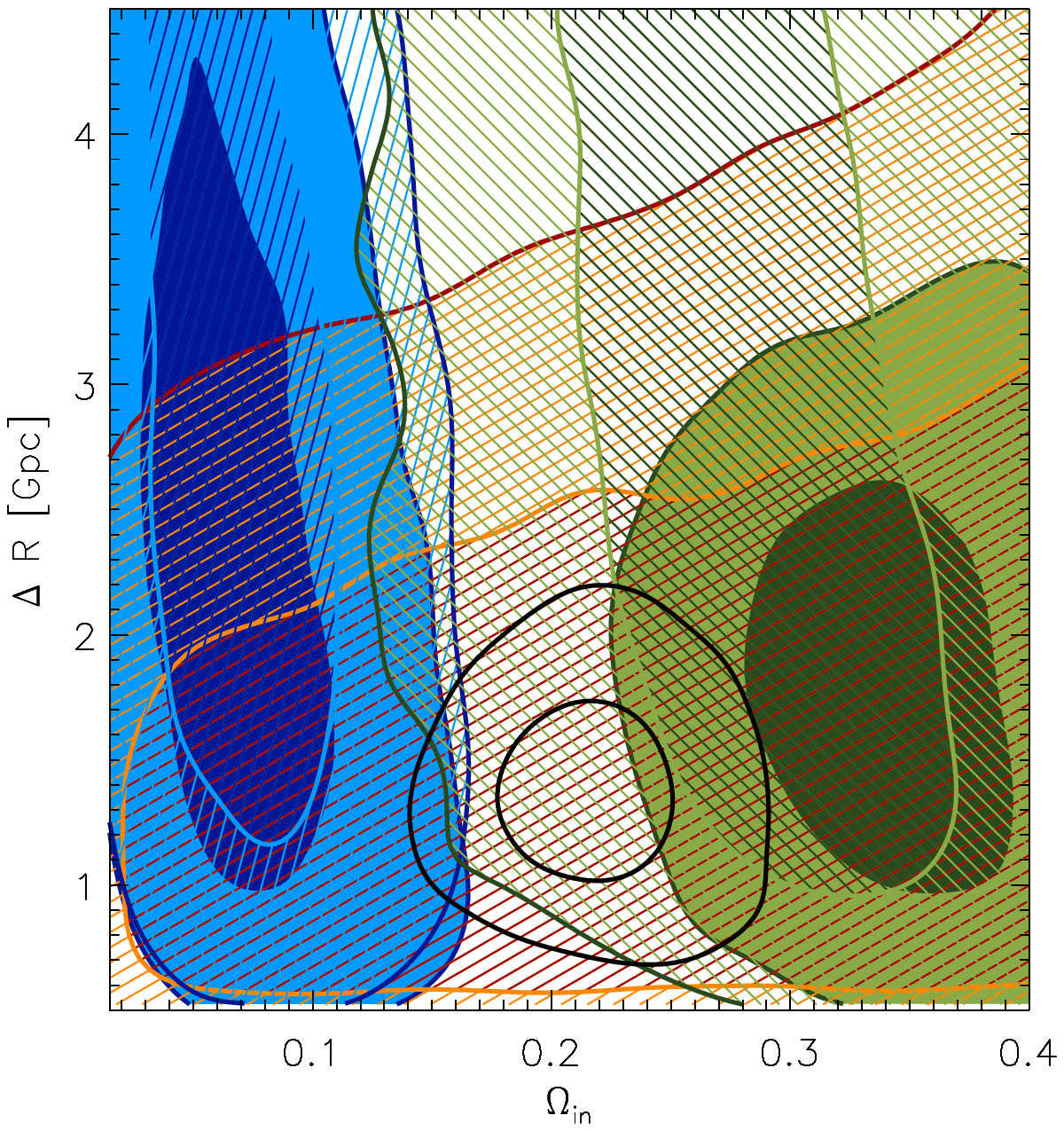}
\includegraphics[width=0.45\columnwidth]{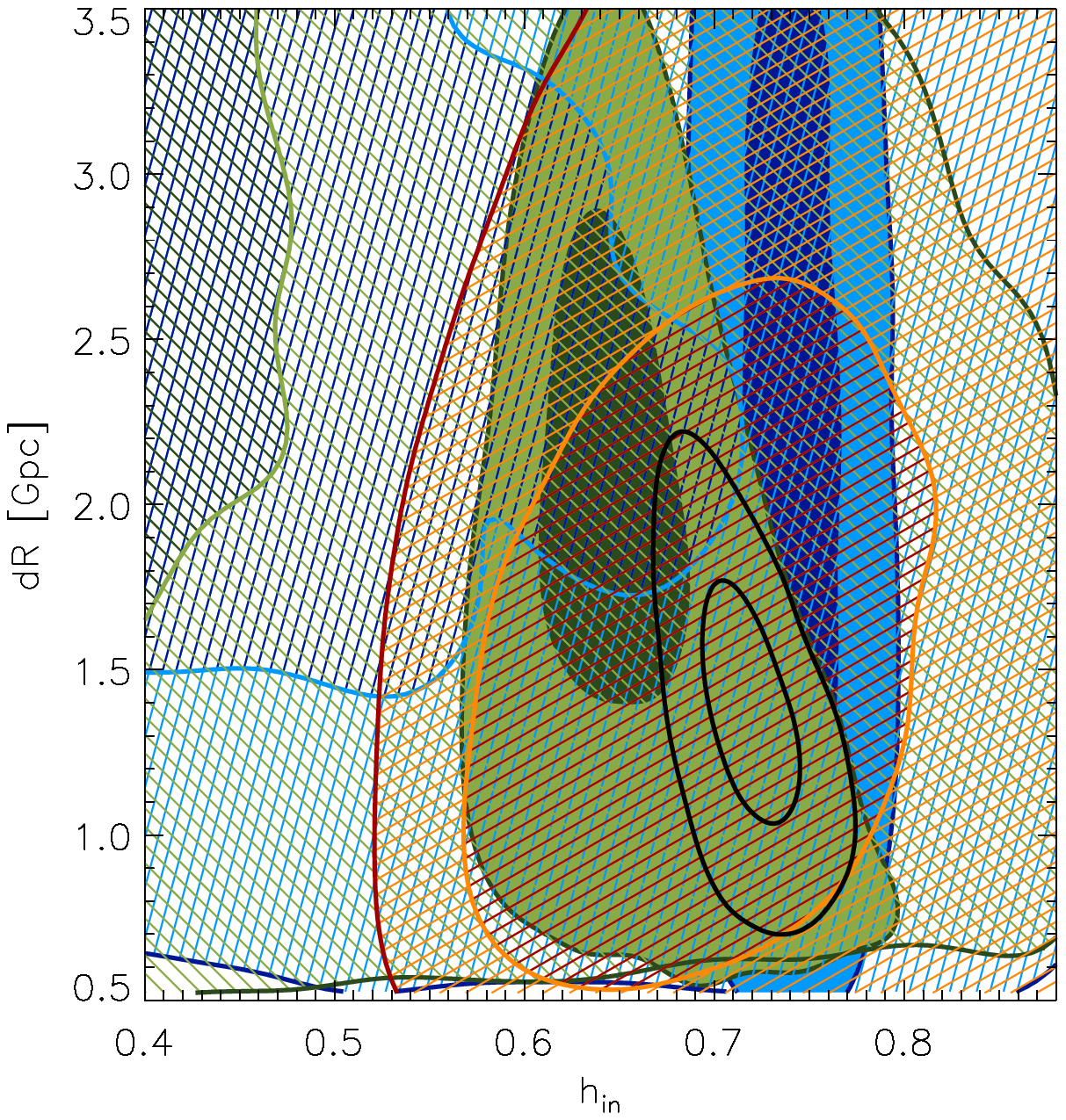} 
\end{center}
\caption{OGBH model. One and two sigma regions for the marginalized likelihood function as obtained from BAO (green), CMB (Orange) and SNe (blue). The filled contours represent the combined constraints using SNe+H0 (blue) and BAO+CMB (green). The black lines correspond to the combined data BAO+CMB+SNe+H0.
\label{contours-ocgbh}} 
\vspace{-1.5cm}
\end{figure}

\begin{figure}[ht!]
\begin{center}
\includegraphics[width=0.45\columnwidth]{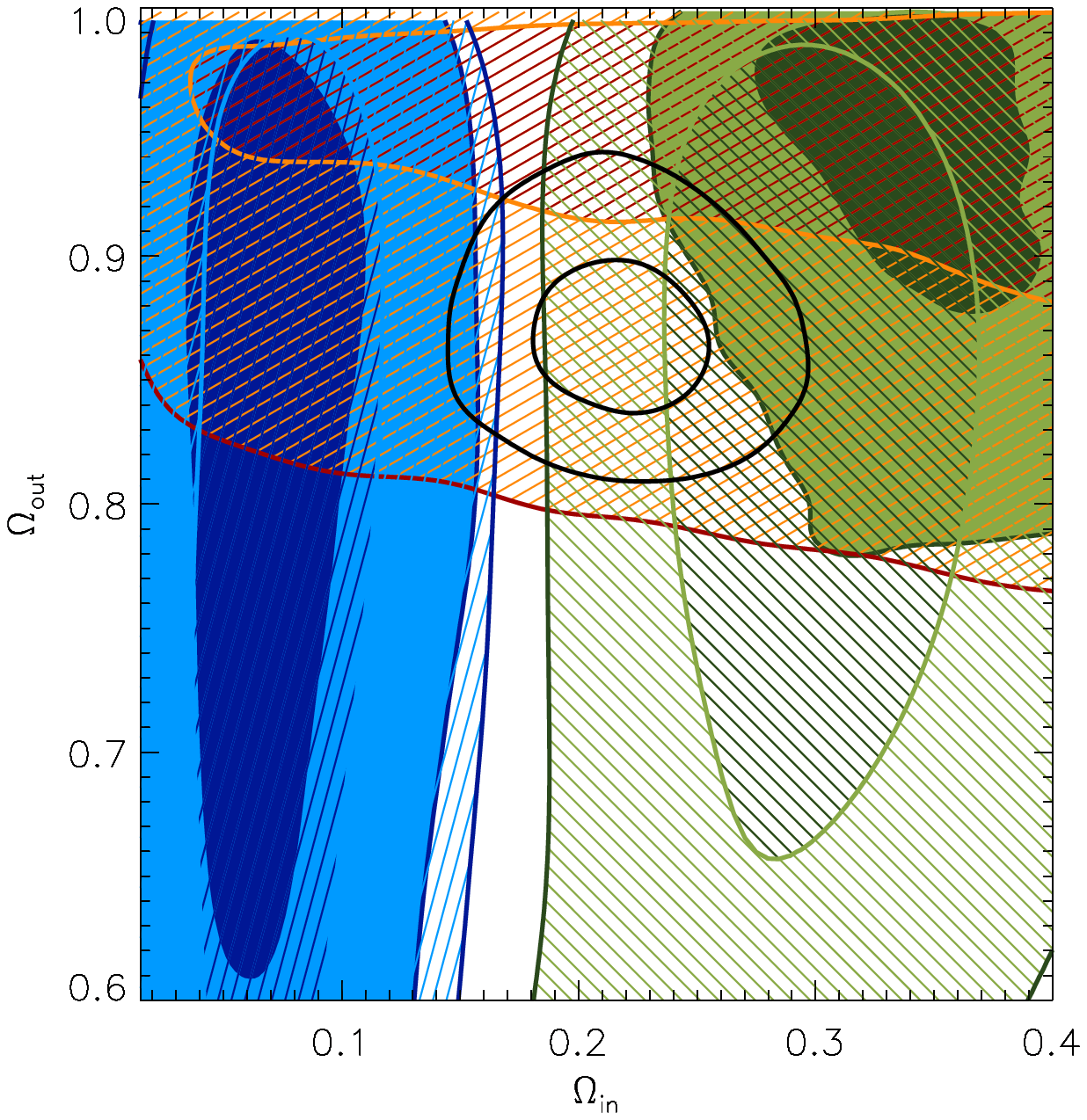}
\includegraphics[width=0.45\columnwidth]{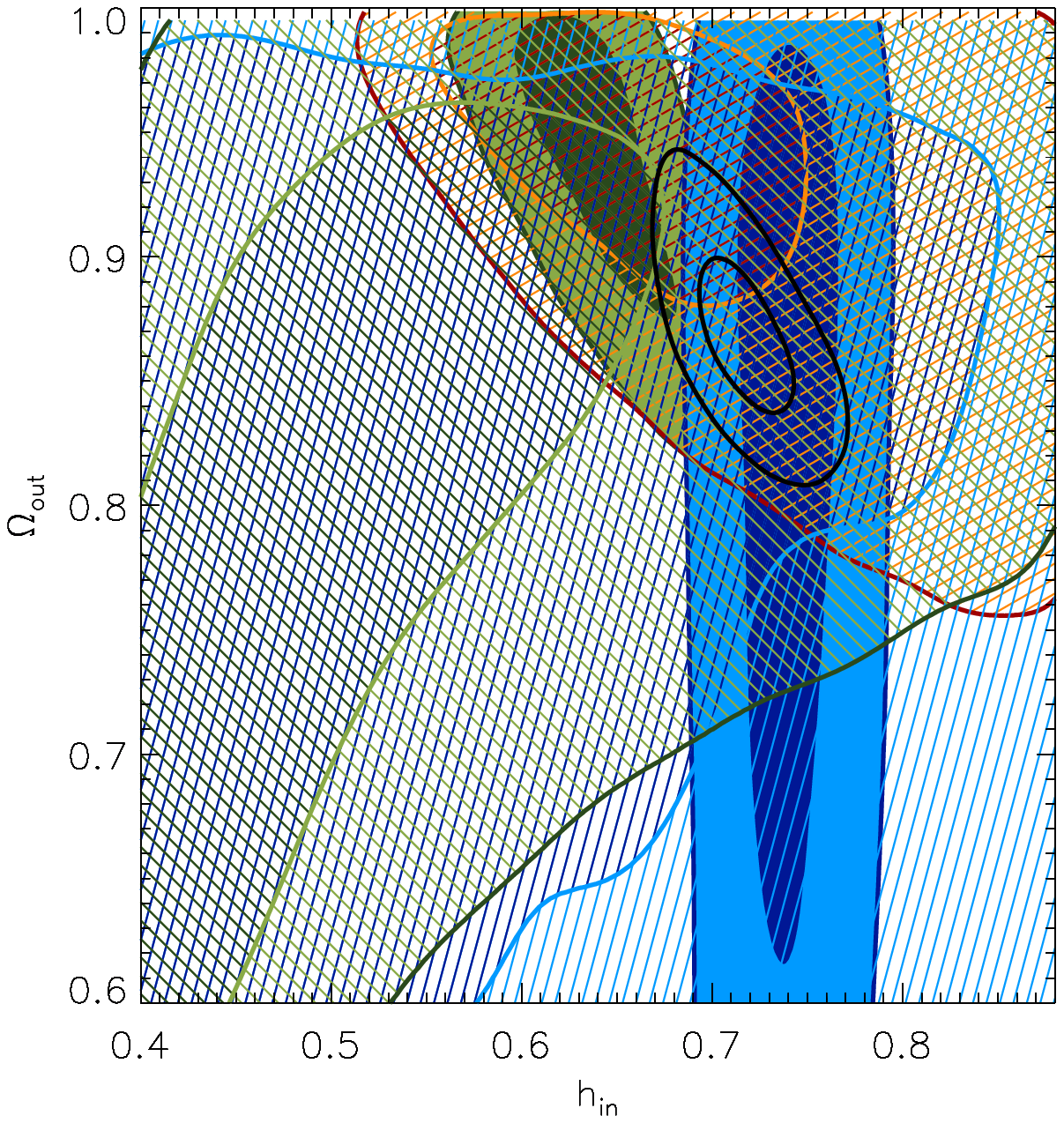} 

\includegraphics[width=0.45\columnwidth]{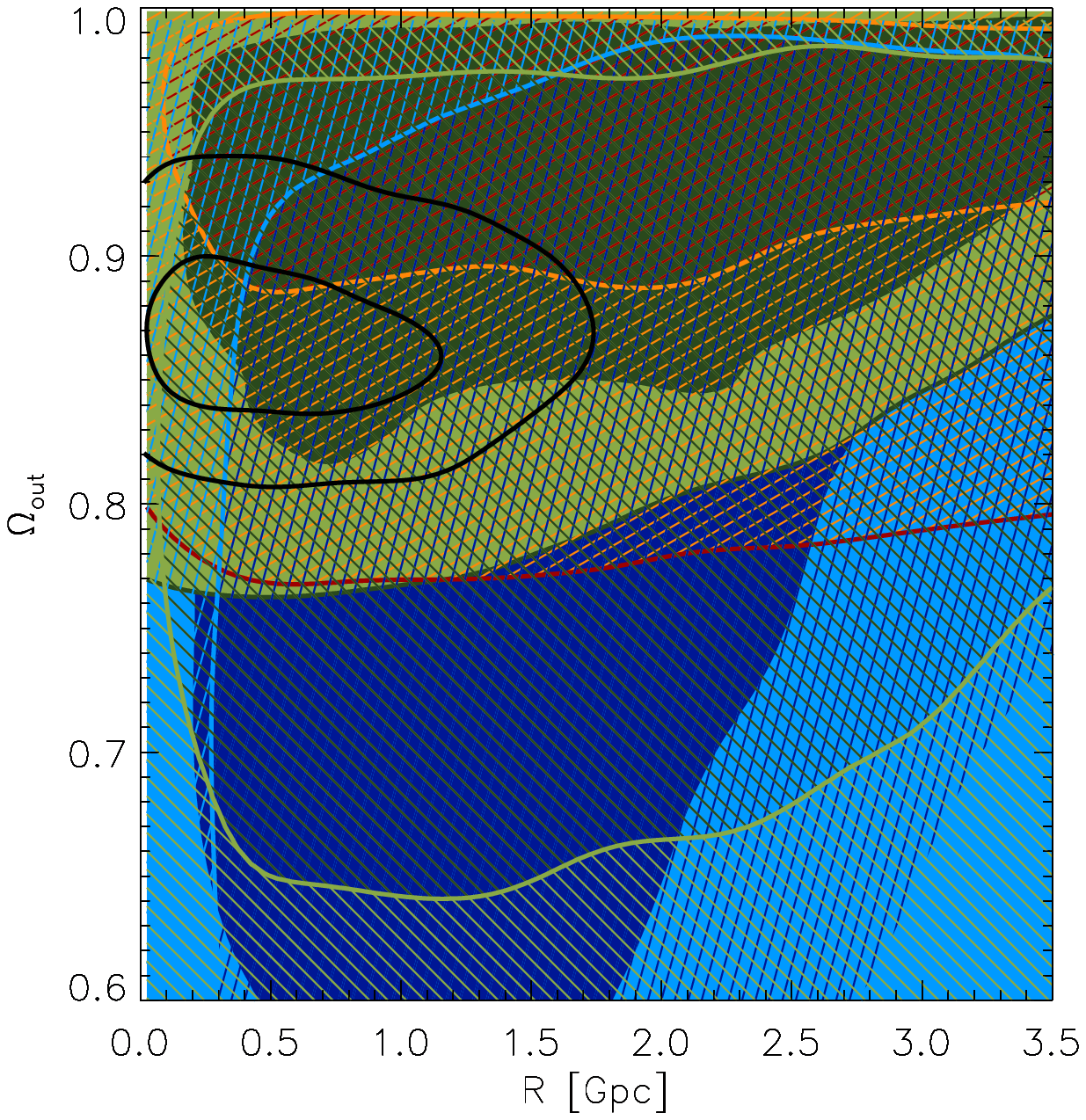} 
\includegraphics[width=0.45\columnwidth]{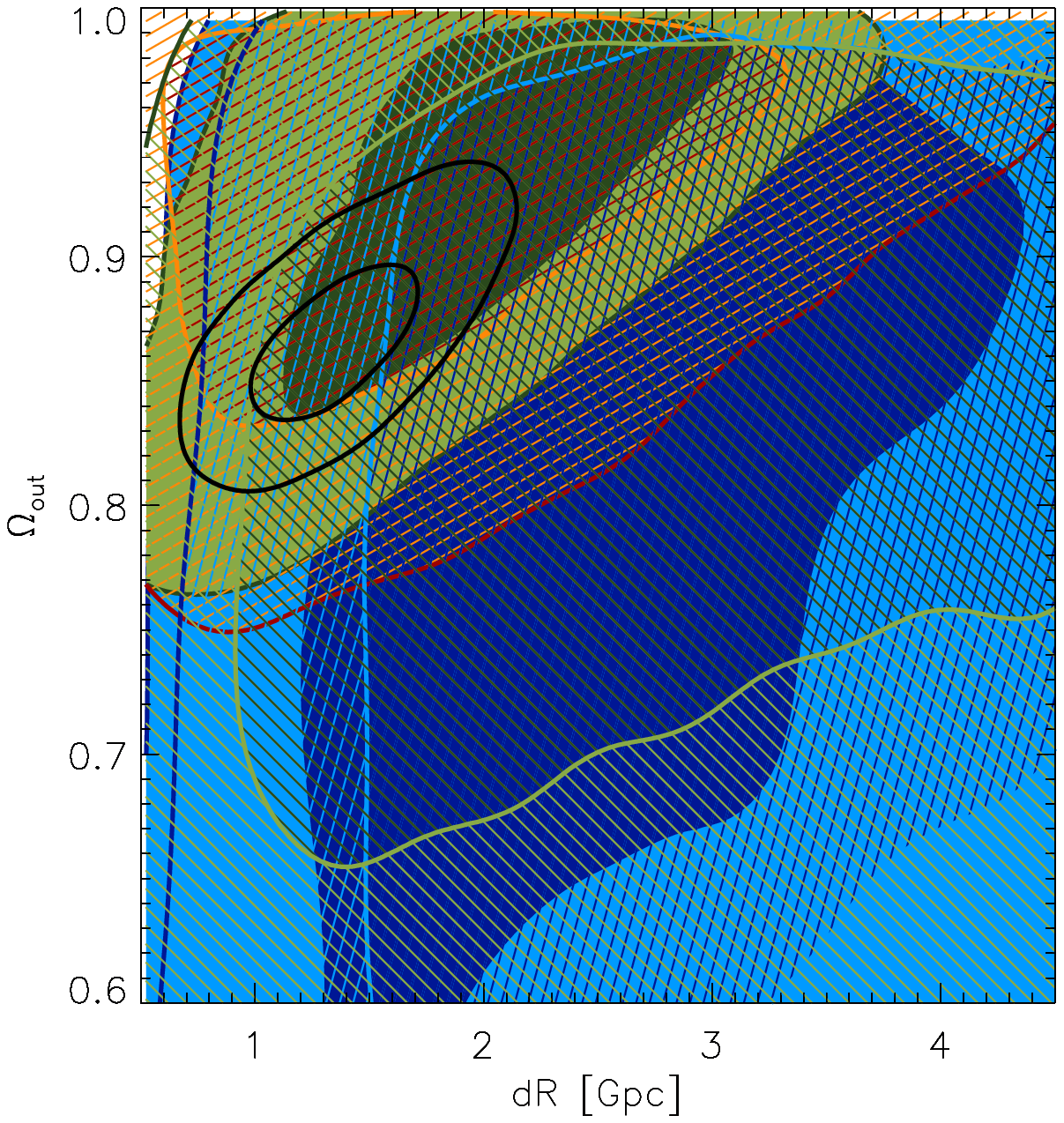}
\end{center}
\caption{Continuation of Figure \ref{contours-ocgbh}.
\label{contours-ocgbh2}} 
\end{figure}

The most severe problem for both models is the existing tension regarding the value of $\Omega_{\rm in}$ determined from BAO and SNe, which differs by roughly $3\sigma$ for the two models (see figures \ref{contours-cgbh}, \ref{contours-ocgbh}). For example, the asymptotically flat model (CGBH) the BAO-only 1D marginalized likelihood yields $ \Omega_{\rm in} = 0.28^{+0.06}_{-0.05} $ (1 sigma), a much higher value than determined by SNe $\Omega_{\rm in} = 0.07\pm 0.04$. This discrepancy is showing how the distance redshift relation necessary to explain the supernovae dimming is incompatible with the stretch of the standard ruler inside the void due to the inhomogeneous rescaling discussed in Section \ref{baoltb}. The low value of $\Omega_{\rm in}$ necessary to fit SNe observations increases the expansion rate and therefore stretches the BAO scale considerably near the center, making it incompatible with the observed values at higher redshift.
This is a purely geometric discrepancy, valid for arbitrary calibration of the standard candles and rulers, and is completely independent of the dynamics originating the characteristic length. Figures \ref{sne}, \ref{bao} show how the best fit models represent a compromise that fails to fit both datasets at low redshift, where the BAO rescaling (Fig. \ref{baorescaling}) is largest.

Naturally, the tension becomes more dramatic when the SNe data are compared to the BAO+CMB combination, because the CMB effectively reduces the allowed range of the initial BAO scale by constraining $H_{\rm in}$, $f_b$ and $\Omega_{\rm out}$. The independence of the constraints on the initial BAO scale is the reason why the asymptotically open model (OCGBH) does not ease the tension between BAO and SNe. This is partly because the apparent freedom gained from allowing $\Omega_{\rm out}$ to vary does not provide essentially different values of the asymptotic BAO scale, already ensured by the freedom in $f_b$. The other reason is that neither SNe nor BAO-only constraints seem to depend on the value of $\Omega_{\rm out}$. We can regard the dashed green and blue contours in Figures \ref{contours-cgbh}, \ref{contours-ocgbh}, \ref{contours-ocgbh2}) as the purely geometric constraints for {\it  arbitrary} values of the standard rulers/candles, while the filled green and blue contours would correspond to adding priors to those calibrations.

\begin{figure}
\begin{center}
\includegraphics[width=0.5\columnwidth]{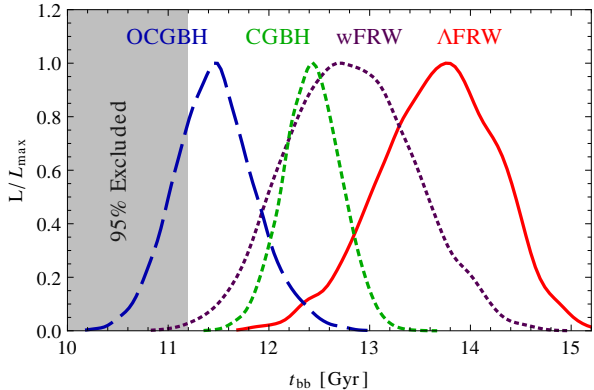} 
\end{center}

\caption{Age of the universe for the best fit models. The curves correspond to marginalization of the MCMC chains over the (homogeneous) Big Bang time (note that it is a derived quantity rather than a parameter varied in the exploration). The gray area shows the region excluded by the age of Globular Clusters in the Milky Way.
\label{age}} 
\end{figure}

The asymptotically flat model (CGBH) also shows a tension between the value of $H_{\rm in}$ determined by H0+SNe and CMB+BAO, both being discrepant at 3$\sigma$. Note that the tension is manifest even using very simplified CMB data, although these yield much looser constraints than the full $C_l$ spectrum. In the asymptotically open model, the additional freedom achieved allows to recover agreement by reducing the value of $\Omega_{\rm out}$, yielding a concordant value for $H_{\rm in}$ from the different datasets. However, the tension would reappear in more a thorough analysis including the full CMB power spectrum data, which typically require $h_{\rm in}\sim 0.4-0.5$ \cite{Moss:2010jx}.
This increase in the local expansion rate also reduces the age of the universe, which is proportional to $H_0(r)^{-1}$. Although it has not been explicitly accounted for in the MCMC, Figure \ref{age} shows how models with a higher expansion rate enter in tension with the limits on the age of the universe obtained from Globular Clusters \cite{Krauss:2003em}, posing yet another difficulty for this type of models.
 
The recovered values of the baryon fraction and the spectral index, mainly determined by the CMB, are much lower than in standard cosmologies. This trend agrees with previous studies in which compensated voids are constrained using the full CMB \cite{Zibin:2008vk}. We note that these features are not relevant to our discussion because 1) the main results are geometric and do not depend on the details of the CMB physics and 2) the obtained values of $f_b,n_s$ rely on a simplified treatment of the CMB peaks.

The best fit models turn out to be rather cuspy, as can be seen in the preference towards $R\approx 0$ Gpc in both the asymptotically flat and open voids. Since the fit is not very good and the individual observations are not very restrictive by themselves (including the combinations BAO+CMB, SNe+H0), this feature might well be due to a compromise between the different datasets, and could be related to the fact that cuspy voids achieve better resemblance to an accelerating universe at low redshift \cite{Vanderveld:2006rb}. In this case, the size of the void is given by the steepness of the inhomogeneity $\Delta R$, which acquires a value $\sim 2.5$ Gpc in the flat case but a smaller value $\sim 1.5$ Gpc in the asymptotically open case. Again, the individual data sets do not yield significant enough information about the value of $R,\Delta R$.
Other than the smallness of the asymptotically open void an the better agreement it gives on the value of $H_0$, there is no significant difference between the two models.

\subsection{Model comparison} \label{modelcompare}

We now proceed to compare the different models under several criteria. The tensions between the different datasets (Figures \ref{contours-cgbh}, \ref{contours-ocgbh} and \ref{contours-ocgbh2}) and the poor fit to SNe and BAO (Figures \ref{sne} and \ref{baodata}) will be reflected in poorer figures with respect to the homogeneous models. Furthermore, inhomogeneous models will be additionally penalized because they have a larger number of parameters.
Table \ref{chi2-table} shows the $\chi^2$ values associated to the different observations and the total values, as well as the result of the different judgment gauges discussed below.

\begin{table}
\begin{center}
\begin{tabular}{| l c c c c  |}
\multicolumn{1}{l}{ }	 & CGBH & OCGBH &  $\Lambda$CDM  & \multicolumn{1}{l}{wCDM}  \\   \hline
Union SNe & \textbf{539.94} 	&\textbf{539.06}  & \textbf{530.70} & \textbf{530.40} \\ 
Hubble $\mu_0$ & \textbf{6.97} 	& \textbf{0.38}   & \textbf{2.17} & \textbf{0.14} \\ \hline

6dF 		& 5.35 		& 4.73 		& 0.35  & 0.09\\
SDSS 		& 0.73 		& 0.04		& 1.29  & 1.24 \\
WiggleZ 	& 0.65 		& 1.20 		& 0.93  & 0.63\\
Carnero et al.	& 0.78		& 0.12 		& 0.61  & 0.34 \\
Total BAO 	
 & \textbf{7.51} & \textbf{6.09}  & \textbf{3.18} & \textbf{2.30}  \\ \hline
Peak positions	& 0.87 	& 0.30 & 0.96  & 0.07 \\
Peak heights	& 1.13 	& 0.11 & 0.24  & 0.04 \\
Total CMB 
& \textbf{2.00} & \textbf{0.41} & \textbf{1.20} & \textbf{0.11} 
\\ \hline 
\textbf{Total $\chi^2$} 
& \textbf{+19.89} & \textbf{+9.40} & \textbf{536.56} & \textbf{-3.62}  \\  \hline 
\# free parameters & 6 & 7 & 5 & 6 \\
$\chi^2/\mbox{d.o.f.}$ & 0.985 & 0.968 & 0.948 & 0.943 \\ \hline
Akaike IC (\ref{aic}) & +22 & +13 & 546.6 & -1.6 \\ 
Bayesian IC  (\ref{bic}) & +26.2 & +15.7 & 568.3 & +2.7 \\ 
Bayes factor (\ref{lnE}) & +10 & +6 & 282.2 & +2.6 \\ \hline
\end{tabular}
\end{center}
\caption{$\chi^2$ contributions to the maximum likelihood models as found by the MCMCs with H0+BAO+CMB+SNe and results from different model comparison criteria discussed in Section \ref{modelcompare}. The values $\chi^2$ as well as the model comparison criteria are given show the total value for the  $\Lambda$CDM best fit, while the other models are given relative to those (minus values are favoured w.r.t. the concordance model, while positive values are disfavoured). The Bayes factor is given by the difference in $-\log E$ (\ref{lnE}).
\label{chi2-table}}
\end{table}

The standard frequentist analysis of parameter estimation, given a set of data, is not very useful for model selection, since it is difficult to compare models with different number of parameters. For instance, the usual method of comparing minimum $\chi^2$ per effective degree of freedom normally misses the point and is not very decisive, as can be seen in the very close values achieved by the different models. 
Other methods to decide which model gives the best description include various Information Criteria which additionally penalize models described by more parameters. Such include the (corrected) Akaike Information Criterion ($AIC$) \cite{Akaike} and Bayesian Information Criterion ($BIC$) \cite{Schwarz}, given by
\begin{eqnarray}
AIC &=& \chi^2_{\rm min} + 2k + {2k(k-1)\over N-k-1}  \,, \label{aic}\\
BIC &=& \chi^2_{\rm min} + k\ln N\,, \label{bic}
\end{eqnarray}
where $k$ is the number of free parameters of a given models and $N$ the number datapoints used in the constraints. The Bayesian evidence
\begin{equation}\label{BE} 
E({\bf D}|{\cal M}) = \int d u\ {\cal L}({\bf D}| u,{\cal M})\, \pi( u,{\cal M})\,,
\end{equation}
is given by the integral of the likelihood ${\cal L}({\bf D}| u,{\cal M})$ over the values of the model parameters $u$ allowed by the priors $\pi( u,{\cal M})$. The computation of the Bayesian evidence is difficult in general, therefore, we will use a simple expression which can be obtained provided that the likelihood is a single isolated peak, far from the edges of the prior ranges \cite{GarciaBellido:2008nz}
\begin{equation}\label{lnE}
-\ln E = -\ln {\cal L}_{\rm max} + \ln A + \sum_i^n \ln ( u_i^{\rm max}- u_i^{\rm min})\,, 
\end{equation}
where $A$ is the normalization of the likelihood, and $[u_i^{\rm min}, u_i^{\rm max}]$ is the range of parameter $u_i$ allowed in the MCMC exploration (assuming a flat priors), $i=1\dots n$. Moreover, for the case of a Gaussian likelihood, 
\begin{equation}
  {\cal L}( u) = A\,\exp\Big[-{1\over2}{\bf u}^{\rm T}C^{-1} {\bf u}\Big]\,,
\end{equation}
we find $A=(2\pi)^{-n/2}/\sqrt{\det C}$, where $C$ is the covariance matrix and $x_i= u_i - \bar u_i$. It is clear that whenever the prior ranges are too big for the likelihood, the Bayesian evidence is penalized. An estimate of the covariance matrix can be obtained assuming that the obtained parameters are independent from each other. In that situation, the covariance matrix is given by the square of the one sigma allowed ranges in each parameter and the determinant becomes $\det C = \prod_i \sigma_i^2$, where we take the average between the upper and lower bounds given in Table \ref{results}. The determinant computed using the eigenvalues of the covariance matrix used in the MCMC sampling yields similar results.

The bottom part of Table \ref{chi2-table} shows similar values for the homogeneous models, but the preferred one depends ultimately on the chosen criterion. It is interesting to note that the Bayes factor favors the simpler $\Lambda$CDM with a difference of 2.6, despite the slightly better $\chi^2$ fit of the wCDM model. This difference is due to the presence of an additional parameter, $w$, together with the large prior postulated for it, $[-5,0]$, relative to the $1\sigma$ region, $\Delta w\approx 0.2$.

The logarithm of the Bayes factor for the LTB models w.r.t. to the fiducial $\Lambda$CDM is 10 and 6, for the asymptotically flat and open GBH inhomogeneous models respectively, due to the bad fit and the larger parameter range explored. Although the asymptotically open model yields a better Bayes factor, even with an additional parameter, both are strongly disfavored according to Jeffreys' scale, since the difference in the logarithm of the Bayes factor is higher than 5. The Akaike and Bayesian Information Criteria also prefer the homogeneous model, the rejection being significantly stronger for the asymptotically flat case due to the extra tension in the local expansion rate. Increasingly accurate data have significantly worsened the fits of inhomogeneous universes, which were compatible just few years ago (c.g. reference \cite{GarciaBellido:2008nz}).

\section{Discussion}

In this paper we presented new constraints on inhomogeneous Lema\^itre-Tolman-Bondi models in the light of the most recent cosmological data, focusing profiles of the GBH type with a space-independent Big Bang and baryon fraction. The inclusion of higher redshift BAO data together with type Ia supernovae allows to reject the models based only on BAO and SNe, independently of other observational data such as the CMB. Additionally, a model independent method to constraint the local expansion rate through a prior on the supernovae luminosity was introduced.

The physical BAO scale at early times was computed in terms of the asymptotic value and then projected to different redshifts using the background LTB metric. This method is justified due to the existence and stability of constant coordinate geodesic solutions, which are expected to be followed by baryonic overdensities in position space. In addition to the time evolution, the BAO scale is shown to become inhomogeneous and anisotropic due to the different expansion rates in the radial and transverse directions. The dependence of the observed BAO scale on both the cosmic distances \emph{and} the evolution of the scale factor leads generically to different predictions than pure distance indicators such as SNe. The departure is largest near the center of the void, precisely because there is less matter to slow down the expansion that drives the growth of the BAO scale. Ultimately, the difference between the two distances can be regarded as a concrete realization of more general tests of the Copernican Principle \cite{Clarkson:2007pz}.

The addition of BAO data at higher redshifts increases considerably their constraining power in this type of models because they help to fix the asymptotic value. The result represents a new drawback for this type of models, as the value of the local matter density $\Omega_{\rm in}\gtrsim 0.2$ preferred by BAO is about 3$\sigma$ apart from the value $\Omega_{\rm in}\lesssim 0.18$ found using Supernovae, as can be inferred from Figures \ref{contours-cgbh} and \ref{contours-ocgbh}. 
The tension between the two datasets persists when asymptotically open models are studied, and worsens when the information from the CMB is added, since it constraints the parameters involved in the acoustic scale determination ($f_b,h,\Omega_{\rm out}$). Asymptotically flat LTB models show an additional tension regarding the value of the local Hubble rate when CMB and BAO are combined. Allowing $\Omega_{\rm out}\leq 1$ relaxes this incompatibility, but we expect it to re-emerge in a more detailed analysis of the CMB. Additionally, larger values of the expansion rate might render the universe too young to account for the ages of stars in globular clusters. The adiabatic GBH models fail to simultaneously fit the data, and a Bayesian analysis shows that they are ruled out at high confidence. 

The above results were obtained for a particular choice of the matter profile. However, the difficulties of the model are manifest in the determined value of matter contrast at the center of the void, while the remaining parameters are poorly constrained by individual datasets. The departure between cosmic rulers and candles becomes most severe at the center of the void, and  we expect that $\Omega_{\rm in}=\lim_{r\to0}\Omega_M(r)$ captures this tension regardless of other features. Since $\Omega_{\rm in}$ can be defined for any LTB model regardless of the parameterization, it is reasonable to expect this result to hold for {\it all} large void models with space-independent Big Bang and baryon fraction. Nonetheless, we have to keep in mind that the SNe and BAO constraints depend on all the parameters through the distance determinations and the evolution of the BAO scale up to a certain redshift, and therefore different shapes for the profile might soften the tension between the two datasets.\footnote{Recently, genetic algorithms were used to analyze the sensitivity of LTB model profiles to the data regardless of the parameterization \cite{Nesseris:2012tt}. The results were not conclusive with respect to the shape of the void profile.}

Similarly to the dependence of CMB constraints with the primordial power spectrum \cite{Nadathur:2010zm}, it is conceivable that fine tuned initial perturbations could be used to reconcile the BAO observations with SNe in adiabatic voids. However, such conditions would not only need to \emph{provide} an enhanced scale to explain the observed feature in galaxy correlation, but also to \emph{hide} the actual BAO scale that would naturally form due to the existence of a preferred scale (the sound horizon at the recombination epoch). On top of this challenging task, the \emph{fake} BAO scale should be shorter near the center to compensate the inhomogeneous growth and fit the observations, therefore requiring some amount of radial dependence that would be at odds with the (quasi) homogeneous initial state.

The effects of the inhomogeneity on the BAO scale are unavoidable. Even if a different void profile might yield a slightly better fit, more precise data e.g. from future surveys such as EUCLID \cite{Amendola:2012ys} will eventually be able to distinguish adiabatic LTB models from the homogeneous case regardless of the shape of the inhomogeneity. Together with the remaining observational problems for large void models with space-independent Big Bang, this sets the stage for abandoning the adiabatic assumption. A scenario with inhomogeneous bang time would require more careful considerations on the origin of the BAO scale to account for the early time inhomogeneity, but it is still possible that the freedom gained from decoupling $H_0(r)$ from $\Omega_{\rm M}(r)$ renders BAO and SNe observations compatible, although the modulation of the Hubble rate is restricted by the local and asymptotic values, fixed by SNe luminosity priors and the CMB.\footnote{Models with arbitrary $H_0(r)$ also predict a too large kinematic Sunyaev-Zel'dovich effect \cite{Bull:2011wi,Zibin:2011ma}. It might be still possible to avoid these constraints by including an additional baryon to photon profile $\eta(r)$ \cite{Clarkson:2012bg}.}

\begin{figure}[t!]
\begin{center}
\includegraphics[width=0.5\columnwidth]{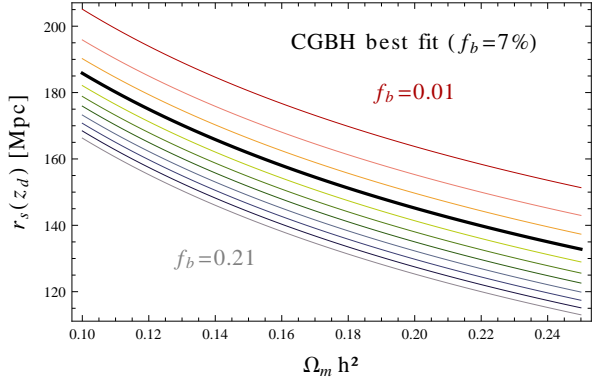} 
\end{center}
\caption{FRW, coordinate baryon acoustic scale $r_s(z_d)$ as a function of the physical matter density $\Omega_m h^2$ for different values of the baryon fraction. The mass of the baryons acts lowering the speed of sound in the baryon-photon fluid, and therefore increasing their amount reduces the resulting acoustic scale, which is related to the sound horizon. An LTB model with a higher baryon fraction near the center might render the BAO and SNe observations compatible by lowering the value of $d_z\propto r_s(z_d)$ near the center of the void (see Figure \ref{bao}). \label{baryons-rs}} 
\end{figure}

A simpler possibility to reconcile SNe and BAO would be to allow for large scale baryon isocurvature modes, and induce a radial dependence on the early time BAO scale through a non-constant baryon to matter ratio $f_b(r)$.\footnote{A related alternative has been explored as way to explain the observed abundances of primordial nuclei and attempt to solve the primordial lithium problem \cite{Regis:2010iq}.} Figure \ref{bao} suggests that lowering the {\it local} value near the center of the void would give a nicer fit to the observations, since the value of $d_z\propto r_s/D_V$ is proportional to the physical acoustic scale. A higher baryon fraction acts reducing the speed of sound of the baryon-photon fluid, therefore shortening the sound horizon that determines the observed BAO scale (see Figure \ref{baryons-rs}). Adding more baryons at the center of the void would provide the necessary freedom to compensate for the inhomogeneous expansion and render the model phenomenologically viable, although more involved and less appealing.\footnote{Baryon isocurvature modes are severely constrained by CMB observations~\cite{Beltran:2005xd,Komatsu:2010fb}. However, these constraints apply to scales much smaller than the size of the void.}

To summarize, we have shown how the BAO scale, acting as a standard (but evolving) ruler and the supernovae explosions, acting as standard candles, lead to different predictions in an inhomogeneous universe, which are disfavored by current data. The conclusion of the analysis is that the use of purely geometric probes, that only recently have become sufficiently constraining, is able to rule out the whole class of adiabatic LTB models. This is independent of other dynamical constraints, like those coming from the kinematic Sunyaev-Zel'dovich effect or the integrated Sachs-Wolfe effect, which in the near future can be used to definitely rule out all inhomogeneous models without dark energy. The present results are also relevant for observationaly constraining more general inhomogeneous models \cite{Bolejko:2011jc} including some recent proposals that also include dark energy \cite{Grande:2011hm,BuenoSanchez:2011wr,Roos:2011bq,Marra:2012pj}.

\acknowledgments

We thank Troels Haugbolle, David Alonso, Savvas Nesseris, Domenico Sapone, Eusebio Sanchez, Enrique Gazta\~naga and Alicia Bueno Belloso for enlightening discussions at various stages of the paper, as well as Sean February, Seshadri Nadathur and James Zibin for correspondence and comments to the first version. We also acknowledge financial support from the Madrid Regional Government (CAM) under the program HEPHACOS S2009/ESP-1473-02, from MICINN under grant  AYA2009-13936-C06-06 and Consolider-Ingenio 2010 PAU (CSD2007-00060), as well as from the European Union Marie Curie Initial Training Network "UNILHC" PITN-GA-2009-237920.
MZ is supported by MICINN (Spain) through the project AYA2006-05369 and the grant BES-2008-009090, and enjoyed an Yggdrasil grant from the Norwegian Research Council while completing this project.

\bibliographystyle{JHEP}

\bibliography{ltbpaper}

\end{document}